\documentclass[reprint,superscriptaddress,
 amsmath,amssymb,
 aps,
pra,
longbibliography
]{revtex4-2}

\usepackage{graphicx}
\usepackage{dcolumn}
\usepackage{bm}
\usepackage{physics}
\usepackage{amsfonts, amsmath}
\usepackage{dsfont}
\usepackage{amsthm}
\usepackage{bm}
\usepackage{bbold}
\usepackage{color}
\usepackage{xcolor}
\usepackage{lipsum,babel}
\usepackage{hyperref}
\usepackage{enumerate}
\usepackage{comment}
\usepackage[normalem]{ulem}
\usepackage[ruled,vlined]{algorithm2e}
\SetKw{Continue}{continue}

\usepackage[utf8]{inputenc}
\usepackage[T1]{fontenc}

\newcommand{\keto}[1]{\vert #1 \rangle}
\newcommand{\brao}[1]{\langle #1 \vert}
\newcommand{\dyado}[1]{\vert #1 \rangle \langle #1 \vert}

\newtheorem{theorem}{Theorem}
\newtheorem{corollary}{Corollary}

\newtheorem{definition}{Definition}
\newtheorem{lemma}{Lemma}

\begin{document}

\definecolor{navy}{RGB}{46,72,102}
\definecolor{pink}{RGB}{219,48,122}
\definecolor{grey}{RGB}{184,184,184}
\definecolor{yellow}{RGB}{255,192,0}
\definecolor{grey1}{RGB}{217,217,217}
\definecolor{grey2}{RGB}{166,166,166}
\definecolor{grey3}{RGB}{89,89,89}
\definecolor{red}{RGB}{255,0,0}

\preprint{APS/123-QED}

\title{Restoring Heisenberg scaling in time via autonomous quantum error correction}
\author{Hyukgun Kwon}
\email{kwon37hg@sejong.ac.kr}
\affiliation{Department of Physics and Astronomy, Sejong University, 209 Neungdong-ro Gwangjin-gu, Seoul 05006, Republic of Korea}
\affiliation{Pritzker School of Molecular Engineering, University of Chicago, Chicago, Illinois 60637, USA}
\affiliation{Center for Quantum Technology, Korea Institute of Science and Technology, Seoul 02792, Republic of Korea}

\author{Uwe R. Fischer}
\email{uwerf@snu.ac.kr}
\affiliation{Seoul National University, Department of Physics and Astronomy, Center for Theoretical Physics, Seoul 08826, Republic of Korea}

\author{Seung-Woo Lee}
\email{swleego@gmail.com}
\affiliation{Department of Physics, Pohang University of Science and Technology(POSTECH), Pohang 37673, Republic of Korea}
\affiliation{Center for Quantum Technology, Korea Institute of Science and Technology, Seoul 02792, Republic of Korea}

\author{Liang Jiang}
\email{liangjiang@uchicago.edu}
\affiliation{Pritzker School of Molecular Engineering, University of Chicago, Chicago, Illinois 60637, USA}

\begin{abstract}
We establish a sufficient condition under which autonomous quantum error correction (AutoQEC) can effectively restore Heisenberg scaling (HS) in quantum metrology. Specifically, we show that if all Lindblad operators associated with the noise commute with the signal Hamiltonian and a particular constrained linear equation admits a solution, then an ancilla-free AutoQEC scheme with finite $R$ (where $R$ represents the ratio between the engineered dissipation rate for AutoQEC and the noise rate,) can approximately preserve HS with desired small additive error $\epsilon > 0$ over any time interval $0 \leq t \leq T$. We emphasize that the error scales as $ \epsilon = O(\kappa T / R^c) $ where $c$ is a positive integer and $\kappa$ is the noise rate, indicating that the required $R$ decreases significantly with increasing $c$ to achieve a desired error. Furthermore, we discuss that if the sufficient condition is not satisfied, logical errors may be induced that cannot be efficiently corrected by the canonical AutoQEC framework. Finally, we numerically verify our analytical results by employing the concrete examples of phase estimation under dephasing noise.
\end{abstract}
              
\maketitle

\section{Introduction}
Quantum metrology investigates how quantum resources can enhance the precision of signal estimation. Over the past years, numerous studies have shown that quantum coherence and entanglement can potentially 
achieve higher estimation precision compared to classical methods
\cite{intro-giovannetti2004quantum,  intro-giovannetti2006quantum, intro-Giovannetti2011, intro-braun2018quantum, intro-Pirandola2018}. 
These advantages have been explored in a wide range of applications, including gravitational wave detection, interferometry, magnetometry, atomic clocks
\cite{intro-caves1981quantum, intro-abadie2011gravitational,  intro-aasi2013enhanced,  intro-PhysRevLett.71.1355,intro-taylor2008high, intro-baumgratz2016quantum, intro-giovannetti2001quantum, intro-komar2014quantum,schneiter}. 
A central goal in quantum metrology is to attain the \textit{Heisenberg scaling} (HS) wherein the variance of the estimator scales as $1/T^{2}$ with total sensing time $T$, representing an ultimate quantum-enhanced precision limit \cite{intro-giovannetti2006quantum}. {However, in practice, achieving HS is often hindered by decoherence induced by interactions with the environment, limiting the attainable precision below HS \cite{noi-demkowicz2009quantum, noi-demkowicz2012elusive, noi-escher2011general, noi-huelga1997improvement}.}

To tackle these challenges, quantum error correction (QEC)--primarily developed within the context of quantum computation--has been employed to restore HS \cite{qec-PhysRevLett.112.080801, qec-PhysRevLett.112.150801, qec-PhysRevLett.112.150802, qec-PhysRevLett.122.040502, qec-PhysRevX.7.041009, qec-zhou2018achieving, qec-PRXQuantum.2.010343,  qec-zhuang2020distributed}. It has been shown that as long as the noise remains distinguishable from the signal--formally characterized by the \textit{Hamiltonian-not-in-Lindblad-span} (HNLS) condition--QEC protocols capable of preserving HS can be constructed \cite{qec-PhysRevX.7.041009, qec-zhou2018achieving, qec-PhysRevLett.122.040502}. 
The necessity of HNLS condition--which does not arise in the context of quantum computation--stems from the fundamental constraint that the signal Hamiltonian is fixed and cannot be arbitrarily manipulated in response to the structure of the noise.
However, even when HNLS condition is satisfied, the QEC protocols capable of restoring HS typically require continuous monitoring and active feed-forward processes (referred to as \textit{ideal QEC} below), which imposes substantial resource demands and presents significant experimental challenges.

\begin{figure*}[t]
    \centering
    \includegraphics[width=\textwidth]{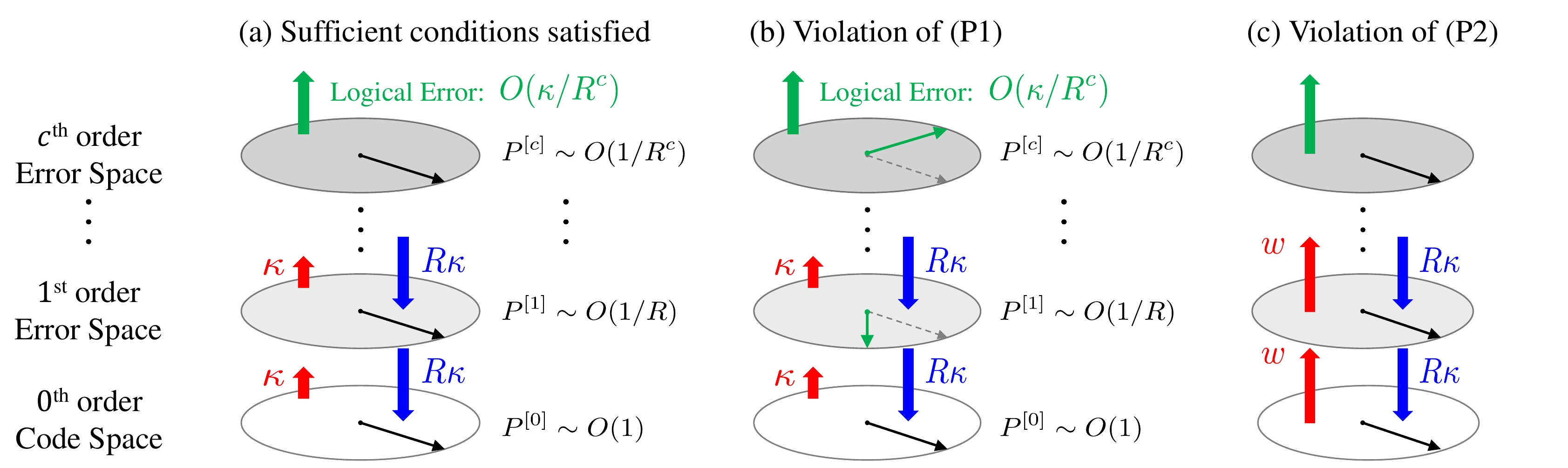}
    \caption{Schematic illustration of AutoQEC dynamics under different conditions. Each panel depicts the evolution of quantum states across the code space and higher-order error spaces. (a) The engineered dissipation (blue thick arrows) effectively counteracts natural dissipation (red thick arrows), leading to a suppression of the probability of occupying the $0\leq n \leq c$-th order error space, scaling as $P^{[n]}\sim O(1/R^{n})$. Consequently, the decoherence (induced by the natural dissipation) rate is suppressed to $O(\kappa/R^{c})$ at the logical level, also relevant to AutoQEC dynamics in quantum computation. (b) Phase differences between subspaces introduce logical errors, represented by the green thin arrows in the error spaces. As a result, even though the decoherence rate is suppressed to $O(\kappa/R^{c})$, one may not achieve AutoQEC up to order $c$. (c) In addition to natural dissipation, $\hat{U}(wt):=e^{-i\hat{H}wt}$ induces additional errors that transfer states into higher-order error spaces, indicated by $w$-induced red thick arrows. These errors accumulate more rapidly than those caused by natural dissipation, in particular
    when $w \gg \kappa$.} 
    \label{fig:AQEC_schematic}
\end{figure*}

Autonomous quantum error correction (AutoQEC) can be a promising candidate to address these challenges. 
AutoQEC leverages engineered dissipation to protect quantum states from noise without continuous monitoring or active feed-forwards, thereby {potentially} avoiding the substantial resource requirements of ideal QEC \cite{aqec-Mirrahimi_2014, aqec-PhysRevLett.116.150501, aqec-PhysRevX.9.041053, aqec-sciadv.aay5901}. By exploiting engineered dissipation, AutoQEC suppresses the decoherence rate $\kappa$ to $O(\kappa/R^{c})$ at the logical level (see Fig.~\ref{fig:AQEC_schematic} (a)), where $R \gg 1$ represents the ratio between the engineered dissipation rate and the decoherence rate, and $c$ is a positive integer referred to as the \textit{order of AutoQEC} that is determined by an AutoQEC code \cite{aqec-lebreuilly2021autonomousquantumerrorcorrection}. In quantum computation, AutoQEC has demonstrated its capability to efficiently suppress errors, both theoretically \cite{aqecth-PhysRevA.98.012317, aqecth-PhysRevLett.111.120501, aqecth-PhysRevLett.120.050503, aqecth-PRXQuantum.3.020302, aqecth-xu2023autonomous, aqecth-albert2019pair} and in experimental implementations \cite{aqecex-gertler2021protecting, aqecex-grimm2020stabilization, aqecex-lescanne2020exponential, aqecex-li2024autonomous}. However, its capability to restore HS remains unclear, as the application of AutoQEC to metrology faces two fundamental challenges: the signal Hamiltonian is fixed, and $R$ is finite in its practical implementation. For instance, Ref.~\cite{qec-rojkov2022bias} explored the application of AutoQEC to quantum metrology; however, both the signal Hamiltonian and the environmental noise introduced logical errors, ultimately preventing the restoration of HS. Furthermore, it is worth emphasizing that, while AutoQEC codes for metrology have been proposed in specific settings \cite{qec-rojkov2022bias, con-reiter2017dissipative}, a general and systematic construction of AutoQEC code tailored for metrology remained unexplored.

In this work, we establish a sufficient condition under which AutoQEC can effectively restore HS in quantum metrology.
Specifically, we show that if all Lindblad operators of the noise commute with the signal and a constrained linear equation (introduced in Theorem \ref{theorem1}) has a solution, then it is possible to construct an ancilla-free AutoQEC code with finite $R$, tailored for quantum metrology that efficiently preserves HS within small additive error $\epsilon>0$ over any given time interval  $0 \leq t \leq T$. (In the remainder of the paper, we refer to this approximate preservation of HS simply as HS.) To the best of our knowledge, this constitutes the first general AutoQEC code explicitly tailored for quantum metrology, and importantly, the code does not require noiseless ancilla. In addition, the error scales as $\epsilon=O(\kappa T / R^{c})$, indicating that, to achieve the targeted error $\epsilon$, the required $R$ decreases substantially with the AutoQEC order $c$. These features can be advantageous for practical implementations, as they both eliminate the need for noiseless ancillas and alleviate the experimental challenge of realizing very large $R$ \cite{aqec-lebreuilly2021autonomousquantumerrorcorrection}.

We also discuss potential challenges arising when the sufficient condition is not satisfied. In such cases, the signal Hamiltonian may induce logical errors that cannot be efficiently corrected by the canonical AutoQEC.  
To validate our results, we apply AutoQEC to two phase-estimation scenarios: (i) a $3$-qubit system governed by the signal Hamiltonian $\hat{H}=\sum_{i=1}^{3}\hat{Z}_{i}$ under correlated dephasing noise, and (ii) a 5-qubit system with $\hat{H}=\prod_{i=1}^{5}\hat{Z}_{i}$ under local dephasing noise. Our numerical results show that when our sufficient condition is satisfied, AutoQEC successfully restores HS.

\vspace*{-0.01em}

\section{Autonomous quantum error correction}
Before introducing our main results, we briefly review AutoQEC in the context of quantum computation \cite{aqec-lebreuilly2021autonomousquantumerrorcorrection}. 
In the presence of time-independent Markovian noise, the dynamics of the system is governed by the Lindblad master equation
\begin{align}
    \frac{d\hat{\rho}(t)}{dt}= -iw[\hat{H},\hat{\rho}(t)]+\kappa\tilde{\mathcal{L}}_{\mathrm{n}}[\hat{\rho}(t)]+R\kappa\tilde{\mathcal{L}}_{\mathrm{E}}[\hat{\rho}(t)]. \label{lindblad}
\end{align}
Here, we have that (1) $\hat{H}$ is the control Hamiltonian for desired logical operations and $w$ is a parameter that governs the corresponding unitary evolution. (2) $\kappa\tilde{\mathcal{L}}_{\mathrm{n}}[\hat{\rho}(t)]:=\sum_{a>0}D[\hat{L}_{\mathrm{n},a}][\hat{\rho}(t)]$ is the natural dissipation which corresponds to noise, with dissipation rate $\kappa$, 
$\hat{L}_{\mathrm{n},a}$ are the Lindblad operators, and $D[\hat{A}][\hat{\rho}(t)]:=\hat{A}\hat{\rho}(t)\hat{A}^{\dagger}-\frac{1}{2}\{\hat{A}^{\dagger}\hat{A},\hat{\rho}(t)\}$. (3) $R\kappa\tilde{\mathcal{L}}_{\mathrm{E}}[\hat{\rho}(t)]:=\sum_{b>0}D[\hat{L}_{\mathrm{E},b}][\hat{\rho}(t)]$ is the engineered dissipation which performs AutoQEC where $R\kappa$ is the engineered dissipation rate, and $\hat{L}_{\mathrm{E},b}$ are the corresponding Lindblad operators. 

The purpose of AutoQEC is to restore the desired target state $\hat{U}_{0}(w t) \hat{\rho}_{0} \hat{U}^{\dagger}_{0}(w t)$ where $\hat{U}_{0}(w t):= e^{-iwt\hat{H}_{0}}$, by applying engineered dissipation $\tilde{\mathcal{L}}_{\mathrm{E}}$ and control Hamiltonian $\hat{H}$. Here, $\hat{\rho}_{0}$ is the initial state that lies within the code space $\mathcal{C}$, and $\hat{H}_{0}$ is the logical Hamiltonian satisfying $\hat{H}_{0}=\hat{\Pi}_{\mathcal{C}}\hat{H}_{0}\hat{\Pi}_{\mathcal{C}}$, where $\hat{\Pi}_{\mathcal{C}}$ is the projection operator onto the code $\mathcal{C}$.

Next, we introduce the definition of \textit{AutoQEC up to order $c$} \cite{aqec-lebreuilly2021autonomousquantumerrorcorrection}:
\begin{definition}\label{definition}
    $\tilde{\mathcal{L}}_{\mathrm{E}}$ and $\hat{H}$ are said to perform AutoQEC up to order $c$ with respect to the code space $\mathcal{C}$, the target logical Hamiltonian $\hat{H}_{0}$ and the natural dissipation $\tilde{\mathcal{L}}_{\mathrm{n}}$, iff there exist two positive constants $M$, $w_{0}>0$ such that for any initial condition $\hat{\rho}_{0}\in \mathcal{C}$ one has
\begin{align}
    \left\lVert   \tilde{\mathcal{P}}_{\mathrm{E}}\left[\hat{\rho}(t)\right]-\hat{U}_{0}(w t) \hat{\rho}_{0} \hat{U}^{\dagger}_{0}(w t) \right\rVert \leq M \left\lVert \hat{\rho}(0) \right\rVert \frac{\kappa t}{R^{c}} \label{aqeccondi}
\end{align}
$\forall~ t,R \geq 0$ and with $w$ satisfying $\abs{w} \leq w_{0}R$. 
\end{definition}
Here, $\tilde{\mathcal{P}}_{\mathrm{E}}:= \lim_{u \to \infty}e^{\tilde{\mathcal{L}}_{\mathrm{E}}u}$ is a CPTP projector stabilizing the code space \cite{aqec-lebreuilly2021autonomousquantumerrorcorrection, aqecCPTP-PhysRevA.89.022118}.
Based on Definition \ref{definition}, we now state 
the main result of Ref.~\cite{aqec-lebreuilly2021autonomousquantumerrorcorrection} (detailed definitions of the relevant terms in the below
lemma are provided further below): 
\begin{lemma}\label{performAQEC}
    If the Knill-Laflamme condition is satisfied for the error set $\mathcal{E}^{[\sim c]}$, then AutoQEC up to order $c$ can be performed by applying the following $\tilde{\mathcal{L}}_{\mathrm{E}}$ and $\hat{H}$:
\begin{align}
    &\tilde{\mathcal{L}}_{\mathrm{E}} = \sum_{n=1}^{c}\sum_{i_{n}=1}^{p_{n}}D[\hat{L}^{[n]}_{\mathrm{E},i_{n}}]+ \sum_{q=1}^{q_{\mathrm{max}}}D[\hat{L}^{[\mathrm{res}]}_{\mathrm{E},q}], \label{engineereddissi}\\
    &\hat{H}=\sum_{j,k=0}^{d_{\mathcal{C}}-1}\sum_{n=0}^{c}\sum_{i_{n}=1}^{p_{n}} \brao{ \mu_{j} } \hat{H}_{0} \keto{ \mu_{k} } \keto{ \mu^{[n]}_{j,i_{n}} } \brao{  \mu^{[n]}_{k,i_{n}} }, \label{AQECHamil} 
\end{align}
where $\hat{L}^{[n]}_{\mathrm{E},i_{n}}=\sum_{j=0}^{d_{\mathcal{C}}-1}\vert \mu_{j} \rangle \langle \mu^{[n]}_{j,i_{n}} \vert ~ \mathrm{  for }~ 1\leq n \leq c ~\mathrm{ and }~ 1\leq i_{n} \leq p_{n}$, and $\hat{L}^{[\mathrm{res}]}_{\mathrm{E},q}= \vert \Phi_{q} \rangle \langle \phi_{q} \vert ~\mathrm{  for } ~1\leq q \leq q_{\mathrm{max}}$. Here, $\keto{\Phi_{q}} $ can be chosen among any normalized states in $\mathcal{C}$.
\end{lemma}
Lemma \ref{performAQEC} contains the followings: 

(1) Error set: Define the zeroth-order error set $\mathcal{E}^{[0]}:=\{\hat{I}\}$, the first-order error set  $\mathcal{E}^{[1]}:=\{\hat{L}_{\mathrm{n},1},\hat{L}_{\mathrm{n},2},\cdots,\hat{L}_{\mathrm{n},N_{\mathrm{n}}}\}$, and the error that corresponds to the no-jump evolution $\hat{B} := \sum_{a=1}^{N_{\mathrm{n}}}\hat{L}^{\dagger}_{\mathrm{n},a}\hat{L}_{\mathrm{n},a}$. We can recursively define the 
$n$-th order error set as 
\begin{align}
    \begin{split}
    \mathcal{E}^{[n]}:= &\{\hat{L}_{\mathrm{n},a}\hat{E}^{[n-1]}_{l} \vert \hat{E}^{[n-1]}_{l} \in \mathcal{E}^{[n-1]}  
    \forall~ a\}\\
    &\cup \{\hat{B}\hat{E}^{[n-2]}_{l} \vert \hat{E}^{[n-2]}_{l} \in \mathcal{E}^{[n-2]} \}.
    \end{split}
\end{align}
Finally, we define the error set up to order $n$ as 
\begin{align}
    \mathcal{E}^{[\sim n]} := \bigcup_{k=0}^{n} \mathcal{E}^{[k]}.
\end{align}

(2) Basis: Let us denote the basis of the code space $\mathcal{C}$ as $\{\keto{\mu_{0}},\keto{\mu_{1}},\cdots , \keto{\mu_{d_{\mathcal{C}}-1}}\}$. By using the Gram-Schmidt algorithm, the following orthonormal basis for the correctable error spaces can be defined as 
\begin{align}
    \begin{split}
    &\mu^{\mathcal{E}^{[\sim c]}}_{i}:=\left\{\keto{\mu_{i}}=\vert {\mu^{[0]}_{i, 1}}\rangle,\vert{\mu^{[1]}_{i, 1}}\rangle,\cdots \vert{\mu^{[1]}_{i, p_{1}}}\rangle,\cdots \vert{\mu^{[c]}_{i, 1}}\rangle,\cdots \vert{\mu^{[c]}_{i, p_{c}}}\rangle\right\} \\
    &= \mathrm{G.S.}\left(\left\{\keto{\mu_{i}},\hat{E}^{[1]}_{1}\keto{\mu_{i}},\cdots \hat{E}^{[c]}_{\abs{\mathcal{E}^{[c]}}}\keto{\mu_{i}}\right\}\right).
    \end{split}\label{gramschmit}
\end{align}
Here, the vectors $\{\keto{\mu^{[n]}_{i,i_{n}}}\}_{i_{n}=1}^{p_{n}}$ represent orthonormal vectors in $n$-th order correctable error space generated from the code word $\keto{\mu_{i}}\in\mathcal{C}$. Furthermore, the residual space is denoted as $\mathcal{R}:= \left(\mathrm{span}\left\{\bigcup_{i=0}^{d_{\mathcal{C}}-1}\mu^{\mathcal{E}^{[\sim c]}}_{i}\right\}\right)^{\perp}=\mathrm{span}\{\keto{\phi_{1}},\keto{\phi_{2}},\cdots , \keto{\phi_{q_{\mathrm{max}}}}\}$ where $q_{\mathrm{max}}=d_{\mathcal{H}}-d_{\mathcal{C}}\sum_{n=0}^{c}p_{n}$, where $d_{\mathcal{H}}$ and $d_{\mathcal{C}}$ is dimension of total Hilbert and code space,  respectively.

\section{Application of AutoQEC to metrology} 
We first clarify the definition of HS.  
Consider a scenario where a parameter $w$ to be estimated is imprinted on a probe state $\hat{\rho}_{0}$ via
unitary $\hat{U}(wt)=e^{-i\hat{H}wt}$, resulting in the ideal signal state $\hat{\rho}_{\mathrm{id}}(t):=\hat{U}(w t) \hat{\rho}_{0} \hat{U}^{\dagger}(w t)$ where $t$ is sensing time. The estimation error of $w$ is fundamentally lower-bounded by the inverse of the quantum Fisher information (QFI) 
$F[\hat{\rho}_{\mathrm{id}}(t)]=4t^{2}\left(\mathrm{Tr}[\hat{H}^{2}\hat{\rho}_{0}]-\mathrm{Tr}[\hat{H}\hat{\rho}_{0}]^{2}\right)$ \cite{est-braunstein1994statistical, est-Helstrom, est-paris2009quantum}. 
Unless the initial probe $\hat{\rho}_{0}$ is an eigenstate of $\hat{H}$, the corresponding QFI scales as $F[\hat{\rho}_{\mathrm{id}}(t)]= \Theta(t^{2})$, which is referred to as HS.

We emphasize that there is a fundamental distinction between the application of AutoQEC in the computation and metrology: In metrology, the signal Hamiltonian $\hat{H}$ is predetermined,  which is thus beyond our control. Consequently, satisfying the Knill-Laflamme condition does not necessarily guarantee AutoQEC up to order $c$ in metrology, as the validity of Lemma \ref{performAQEC} depends on the specific design of the Hamiltonian. This distinction underscores unique challenges of applying AutoQEC in metrological scenarios and suggests that a stricter condition than the Knill-Laflamme criterion may be required. Motivated by this, we investigate conditions under which the application of AutoQEC restores HS. 

Before presenting our sufficient condition, we introduce matrices that simplify the description of our results. Let us express the Hamiltonian $\hat{H}$ in diagonal form as
\begin{align}
    \hat{H}=\sum_{i=0}^{d-1}h_{i}\left(\sum_{l=1}^{N_{i}}\dyado{h^{(l)}_{i}}\right),
\end{align}
where $\{h_{i}\}_{i=0}^{d-1}$ are distinct eigenvalues, and $\{\keto{h^{(k)}_{i}}\}_{k=1}^{N_{i}}$ the corresponding eigenvectors. Next, we define the following operator set:
\begin{align}
    \mathcal{K}^{[\sim c]}:=\{\hat{E}^{\dagger}_{a}\hat{E}_{b} \vert \hat{E}_{a},\hat{E}_{b} \in \mathcal{E}^{[\sim c]}~\forall~ a,b \}.\label{formethod1}
\end{align}
Notably, the Knill-Laflamme condition can be expressed as 
\begin{align}
    \brao{\mu_{i}}\hat{K}_{k}\keto{\mu_{j}}=\sigma_{k}\delta_{ij}~\forall~ 1 \leq k \leq \abs{\mathcal{K}^{[\sim c]}},    \label{formethod2}
\end{align}
where $\hat{K}_{k} \in \mathcal{K}^{[\sim c]}$ for all $k$, and $\sigma_{k}$ are constants associated with $\hat{K}_{k}$.
Next, using $\mathcal{K}^{[\sim c]}$ and the eigenvectors of $\hat{H}$, we define $\abs{\mathcal{K}^{[\sim c]}} \times N_{i}$ matrix $\mathbf{A}^{[\sim c]}_{i}$, with elements 
\begin{align}
[\mathbf{A}^{[\sim c]}_{i}]_{kl} :=\brao{h^{(l)}_{i}} \hat{K}_{k} \keto{h^{(l)}_{i}}~\forall~ k,l. \label{formethod3}
\end{align}
Based on $\mathbf{A}^{[\sim c]}_{i}$, we introduce the following Theorem:
\begin{theorem} \label{theorem1}
    Suppose the following two conditions hold: (T1) $[\hat{H},\hat{L}_{\mathrm{n},a}]=0~ \forall~ a$, and (T2) $\exists i,j \neq i$, along with two probability vectors $\vb*{p}_{i} \in \mathbb{R}^{N_{i}}_{\geq 0}$, $\vb*{p}_{j} \in \mathbb{R}^{N_{j}}_{\geq 0}$ such that $\mathbf{A}^{[\sim c]}_{i}\cdot \vb*{p}_{i}=\mathbf{A}^{[\sim c]}_{j}\cdot \vb*{p}_{j}$.
    
    Then, for any given time $T \geq 0$, and an arbitrarily small error $\epsilon>0$, there exists an ancilla-free AutoQEC scheme with a finite $R$ such that 
    \begin{align}
    \forall~ 0\leq t \leq T:~F[\hat{\rho}(t)] \geq (h_{i}-h_{j})^{2}t^{2} - \epsilon, \label{eqtheorem}
    \end{align}
    where $\hat{\rho}(t)$ is the AutoQEC state, and $F[\hat{\rho}]$ is the QFI of $\hat{\rho}$ with respect to $w$. The relationship between $R$, $\kappa$, $T$, and $\epsilon$ here is given by $\epsilon=O(\kappa T / R^{c})$.
\end{theorem}

{\it Proof sketch [see {Method or} Supplemental Material (SM) Sec. S1 for details].} 
We consider the following two-dimensional code $\mathcal{C}$ where each codeword is composed of linear combination of eigenvectors of $\hat{H}$ corresponding to the same eigenvalue:  
\begin{align}
    \ket{\mu_{0}}:=\sum_{k=1}^{N_{i}}{\sqrt{[\vb*{p}_{i}]_{k}}}\keto{h^{(k)}_{i}},~~\ket{\mu_{1}}:=\sum_{k=1}^{N_{j}}{\sqrt{[\vb*{p}_{j}]_{k}}}\keto{h^{(k)}_{j}}. \label{codewords1}
\end{align}
First, we demonstrate that for this code, {(T1) guarantees $\brao{\mu_{0}}\hat{K}_{k}\keto{\mu_{1}}=0$ without the need for any ancilla, thereby aligning with previous ancilla-free ideal QEC results \cite{qec-PhysRevLett.122.040502}} and (T2) ensures that there exist $\vb*{p}_{i}$ and $\vb*{p}_{j \neq i}$ such that $\brao{\mu_{0}}\hat{K}_{k}\keto{\mu_{0}}=\brao{\mu_{1}}\hat{K}_{k}\keto{\mu_{1}}~\forall~ \hat{K}_{k} \in \mathcal{K}^{[\sim c]}$, i.e., the Knill-Laflamme condition holds for the error set $\mathcal{E}^{[\sim c]}$. 

{
Next, we consider the initial probe state 
\begin{align}
    \hat{\rho}_{0}:=\dyad{\psi_{0}},~~\ket{\psi_{0}}:=\frac{\ket{\mu_{0}}+\ket{\mu_{1}}}{\sqrt{2}}.
\end{align}
For the noiseless scenario, $\hat{\rho}_{0}$ achieves HS, yielding the corresponding QFI $F[\hat{\rho}_{\mathrm{id}}(t)] = (h_i - h_j)^2 t^2$ where the imprinted state is given by
\begin{align}
    \hat{\rho}_{\mathrm{id}}(t):=e^{-i\hat{H}wt}\hat{\rho}_{0}e^{i\hat{H}wt}.
\end{align}
Since $\ket{\mu_{0}}$ and $\ket{\mu_{1}}$ are eigenstates of $\hat{H}$, the evolution can be equivalently expressed as
\begin{align}
    \hat{\rho}_{\mathrm{id}}(t)=e^{-i\hat{H}_{0}wt}\hat{\rho}_{0}e^{i\hat{H}_{0}wt},
\end{align}
with the effective (logical) Hamiltonian
\begin{align}
    \hat{H}_{0}:= h_{i}\dyad{\mu_{0}}+h_{j}\dyad{\mu_{1}}.
\end{align}}

We then establish that AutoQEC up to order $c$, as defined in Eq.~\eqref{aqeccondi}, is achievable. Moreover, Eq.~\eqref{aqeccondi} implies 
\begin{align}
    \abs{F[\tilde{\mathcal{P}}_{\mathrm{E}}\left[\hat{\rho}(t)\right]]-F[\hat{\rho}_{\mathrm{id}}(t)]} \leq \epsilon. \label{qfidifference}
\end{align}
Here, since $\tilde{\mathcal{P}}_{\mathrm{E}}$ is a CPTP map, the data processing inequality of the QFI ensures \cite{est-petz2007quantum}
\begin{align}
    F[\tilde{\mathcal{P}}_{\mathrm{E}}\left[\hat{\rho}(t)\right]] \leq F[\hat{\rho}(t)] \leq F[\hat{\rho}_{\mathrm{id}}(t)]. \label{qfiinequality2}
\end{align}
{Combining Eqs.~\eqref{qfidifference} and \eqref{qfiinequality2} then yields Eq.~\eqref{eqtheorem}. Finally, we emphasize that the final CPTP map $\tilde{\mathcal{P}}_{\mathrm{E}}$ is not required to restore HS. Indeed, by the data-processing inequality in Eq.~\eqref{qfiinequality2}, any such CPTP map does not increase the QFI. Therefore, it is unnecessary to apply $\tilde{\mathcal{P}}_{\mathrm{E}}$ to restore HS by the AutoQEC scheme.}
\qed

{We highlight several important aspects of Theorem \ref{theorem1}. First, our code does not require any noiseless ancilla. Second, the error scales as $\epsilon = O(\kappa T / R^{c})$, implying that increasing the AutoQEC order $c$ enables a fixed target accuracy $\epsilon$ to be achieved with a smaller engineered dissipation strength $R$. This is particularly advantageous in practical regimes where achieving large $R$ is subject to stringent constraints. At the same time, it is important to emphasize that achieving higher-order AutoQEC generally requires a more intricate recovery structure; in the canonical construction, this manifests as an increased number of recovery Lindblad operators $\hat{L}^{[n]}_{\mathrm{E},i_{n}}$, as indicated by Eq. \eqref{engineereddissi} [see SM Sec. S5 for details]. However, rather than constituting a limitation, this reflects a favorable and systematic trade-off: higher-order AutoQEC reduces the required dissipation strength while affording additional flexibility in the recovery design. This interplay between dissipation resources and recovery complexity provides a practical pathway toward efficient implementations.}

Third, identifying $(\vb*{p}_{i}$, $\vb*{p}_{j})$, which corresponds to constructing an AutoQEC code tailored for metrology, reduces reduces to solving a constrained linear feasibility problem. {This can be addressed in a systematic manner using \textit{linear programming} methods, and can be efficiently performed particularly for modest system sizes [see SM Sec. S2].}  
Third, if our sufficient condition is satisfied, HNLS condition which is the necessary and sufficient condition for ideal QEC to fully restore HS \cite{qec-zhou2018achieving}, is automatically satisfied [see SM Sec. S1]. This follows from the fact that the code defined in Eq. \eqref{codewords1} satisfies the Knill-Laflamme condition and ensures $\hat{\Pi}_{\mathcal{C}}\hat{H}\hat{\Pi}_{\mathcal{C}} \neq h\hat{\Pi}_{\mathcal{C}}$ where $h$ is a constant, which implies that HNLS is satisfied \cite{qec-zhou2018achieving, HNLS}.

Finally, we discuss the scenarios where the sufficient condition in Theorem \ref{theorem1} is not satisfied. In Theorem 2 of SM Sec. S3. A, we show that even if the sufficient condition does not hold, AutoQEC can restore HS in the infinite $R$ limit and with the use of a noiseless ancilla, provided that HNLS condition is satisfied. However, for finite $R$, it is not guaranteed that an AutoQEC scheme can efficiently restore HS without the satisfaction of the sufficient condition. To gain further insight into this limitation, we now explore cases where the sufficient condition is not satisfied. When the sufficient condition is indeed satisfied and exploiting the code defined in Eq. \eqref{codewords1}, the signal Hamiltonian $\hat{H}$ exhibits the following two significant properties: 
{
\begin{enumerate}
    \item[(P1)] $\hat{H}$ commutes with all the errors, i.e., $[\hat{H},\hat{E}]=0~\forall~ \hat{E} \in \mathcal{E}^{[\sim c]}$.
    \item[(P2)] $[\hat{\Pi}^{[n]},\hat{H}]=0~\forall~ n$,
where $\hat{\Pi}^{[n]}:=\sum_{i=0}^{1}\sum_{i_{n}}\dyado{\mu^{[n]}_{i,i_{n}}}$ are projection operators that correspond to an $n$-th order correctable error space with $0 \leq n \leq c$.
\end{enumerate}
Here, (P1) follows directly from (T1). To show that (P2) is guaranteed under conditions (T1) and (T2), we assume that both conditions are satisfied. We first observe from Eq.~\eqref{codewords1} that $\hat{\Pi}^{[0]}$ commutes with $\hat{H}$, since each codeword $\ket{\mu_{0}}$ and $\ket{\mu_{1}}$ is composed of the eigenvectors of $\hat{H}$ with same eigenvalue. Furthermore, by construction in Eq. \eqref{gramschmit}, the orthonormal basis for the $n$-th order correctable error spaces  $\{\keto{\mu^{[n]}_{i,i_{n}}}\}_{i_{n}=1}^{p_{n}}$ consists with the linear combinations of
\begin{align}
    \left\{\keto{\mu_{i}},\hat{E}^{[1]}_{1}\keto{\mu_{i}},\cdots \hat{E}^{[n]}_{\abs{\mathcal{E}^{[n]}}}\keto{\mu_{i}}\right\}.
\end{align}
Combined with the fact that $[\hat{H},\hat{\Pi}^{[0]}]=0$, this directly guarantees (P2), i.e., $[\hat{H},\hat{\Pi}^{[n]}]=0$.}
As a consequence, violation of the sufficient condition can lead to violation of either (P1) or (P2), and we now address potential challenges. To this end, we consider the engineered dissipation in Eq. \eqref{engineereddissi}, a canonical engineered dissipation in AutoQEC. First, consider the case when (P1) is violated; $\hat{U}(wt)$ then possibly acts differently across 
code and error spaces, resulting in different phase accumulations between these spaces. These phase differences may induce logical errors that cannot be corrected by engineered dissipation [see Fig.~\ref{fig:AQEC_schematic} (b)]. Second, consider the case when (P2) is violated. In this scenario, $\hat{U}(wt)=e^{-i\hat{H}wt}$  may introduce additional errors that transfer a quantum state to higher-order error spaces. This becomes particularly critical when $w \gg \kappa$ (which is a case of interest in metrology), since then these errors accumulate significantly faster than those arising from natural dissipation [see Fig.~\ref{fig:AQEC_schematic} (c)]. Moreover, it is not guaranteed that engineered dissipation can correct such \textit{coherent} errors. Consequently, if our sufficient condition is not satisfied, AutoQEC does not guarantee efficient restoration of HS. More specifically, considering Theorem 2 in SM, if our sufficient condition is violated we may require substantially larger $R$ to achieve the same error $\epsilon$ compared to the case when the sufficient condition is satisfied. In Sec.~S3. B of the SM, we numerically implement these scenarios. 

\section{Numerical results} 
For the numerical simulation of our proposed scheme, 
we consider $N$-qubit phase estimation in the presence of the correlated dephasing noise. The sensing dynamics (in the absence of AutoQEC) is governed by the Lindbladian evolution 
\begin{align}
    \frac{d\hat{\rho}}{dt}= -iw[\hat{H},\hat{\rho}(t)]+\kappa \sum_{i,j=1}^{N}[\mathbf{C}]_{ij}\left(\hat{Z}_{i}\hat{\rho}\hat{Z}_{j}-\frac{1}{2}\{\hat{Z}_{i}\hat{Z}_{j},\hat{\rho}\}\right), \label{lindnume}
\end{align}
where $\hat{Z}_{i}$ represents the Pauli-$Z$ operator acting on the $i$th qubit and $\mathbf{C}$ is the \textit{correlation matrix} that characterizes the spatial structure of the noise. Specifically, the off-diagonal elements $[\mathbf{C}]_{ij}$ quantify the noise correlation between $i$th qubit and $j$th qubit. 

\begin{figure}[t]
    \centering
    \includegraphics[width=\linewidth]{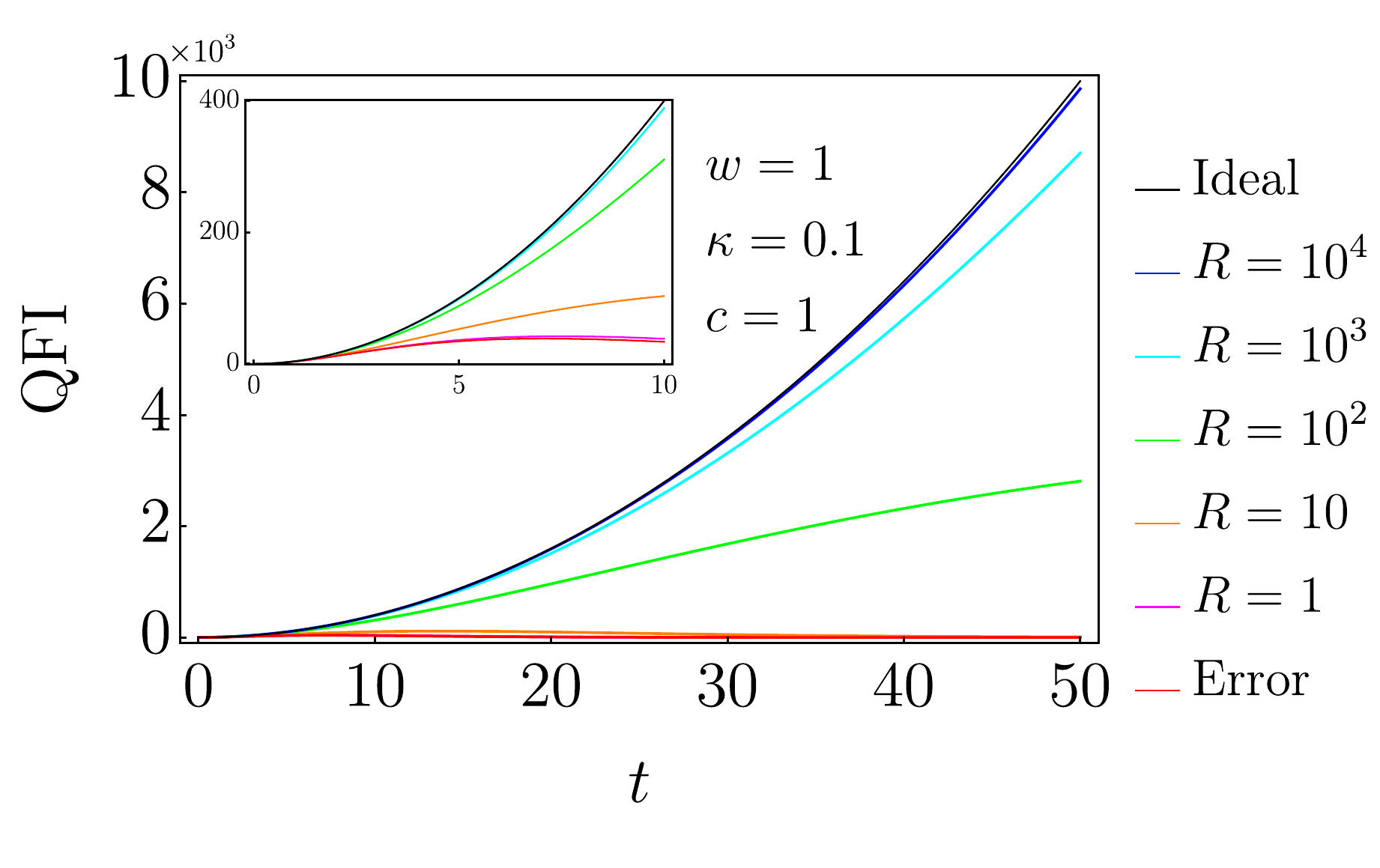}
    \caption{QFI as a function of sensing time $t$ when the signal Hamiltonian is given by $\hat{H}=\sum_{i=1}^{3}\hat{Z}_{i}$ in the presence of correlated dephasing noise defined in Eq. \eqref{correlationmatrix}. 
   With the code defined in Eqs. \eqref{correcodewords1} and \eqref{correcodewords2}, AutoQEC of order $c=1$ can be achieved. In the numerics, we consider  various values of $R$, with $w=1$, $\kappa=0.1$, and AutoQEC order $c=1$. The black line (labeled ``Ideal") and the red line (labeled ``Error") represent the QFI of the noiseless case and without AutoQEC case respectively. The inset shows a zoomed-in view of the QFI at smaller times $t$.
    }
    \label{fig:correlatedepol}
\end{figure}

In the first numerical analysis, we consider the signal Hamiltonian $\hat{H}=\sum_{i=1}^{N}\hat{Z}_{i}$. We note that in this setting, the ideal QEC has been shown to restore HS {as long as a correlation matrix is not full-rank; equivalently, HNLS condition is satisfied} \cite{qec-zhou2018achieving, qec-PhysRevLett.122.040502}. However, for AutoQEC, {a non-full-rank correlation matrix does not always guarantee fulfilling the conditions of} Theorem \ref{theorem1} (see SM 
Sec. S1.A for details). 
We consider a $3$-qubit system with $w=1$, $\kappa=0.1$, and 
\begin{align}
    \mathbf{C}=\frac{1}{10}\begin{pmatrix}
       16 & -4 & -4\\
        -4 & 7 & -5 \\
        -4 & -5 & 7 \\
    \end{pmatrix}. \label{correlationmatrix}
\end{align}
We note that we carefully chose a correlation matrix that satisfies our sufficient condition. Now the sufficient condition in Theorem \ref{theorem1} is satisfied up to AutoQEC order $c=1$. Therefore, based on 
Eq.~\eqref{codewords1}, we consider the following code words:
\begin{align}
    &\keto{\mu_{0}}=(\sqrt{4}\keto{100}+\sqrt{3}\keto{010}+\sqrt{3}\keto{001})/\sqrt{10}, \label{correcodewords1}\\
    &\keto{\mu_{1}}=(\sqrt{4}\keto{011}+\sqrt{3}\keto{101}+\sqrt{3}\keto{110})/\sqrt{10}.\label{correcodewords2}
\end{align}
In Fig.~\ref{fig:correlatedepol}, we show that HS can be efficiently restored by the AutoQEC. Especially, as $R$ increases, HS is more effectively restored.

\begin{figure}[t]
    \centering
    \includegraphics[width=\linewidth]{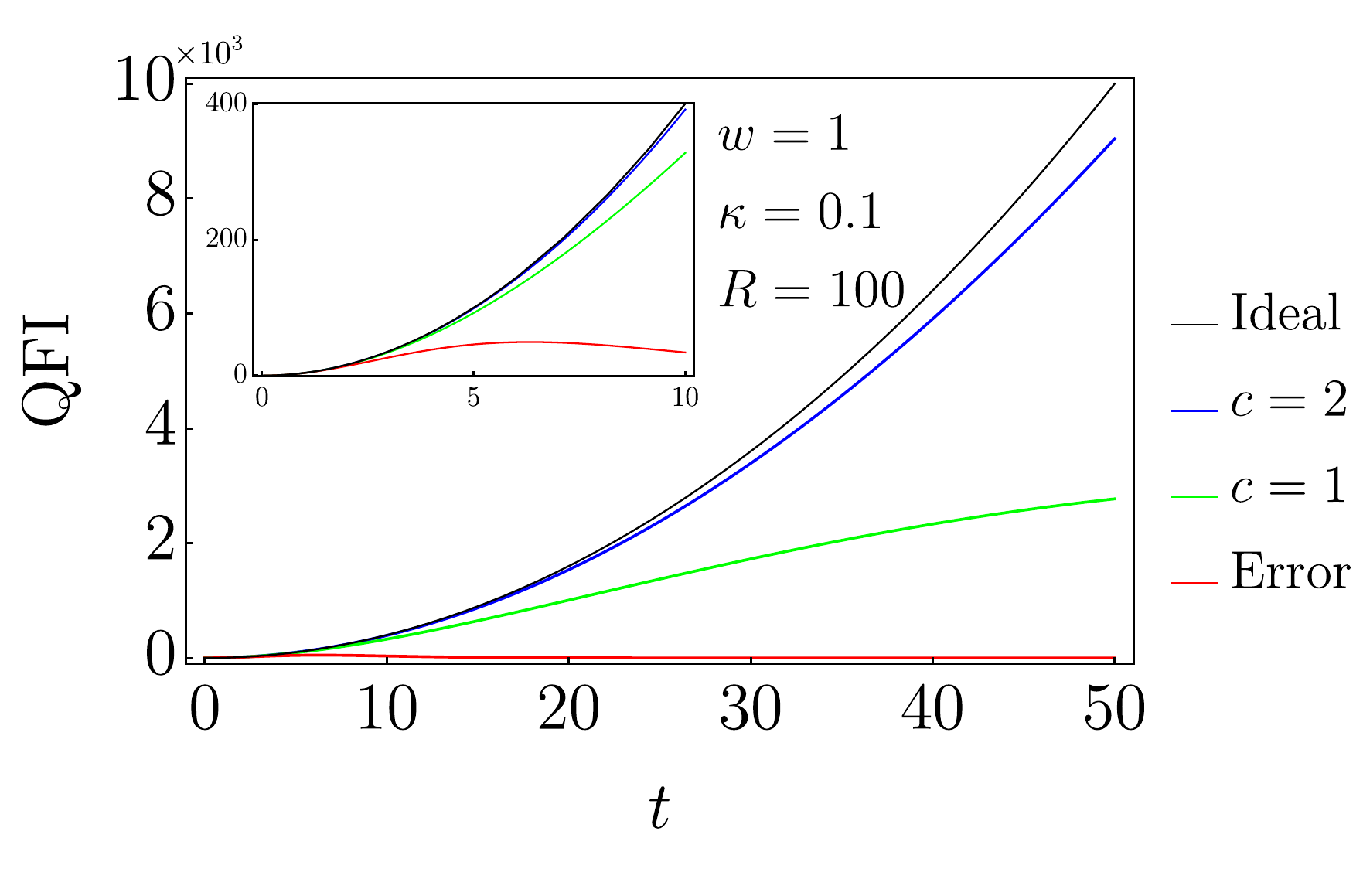}
    \caption{QFI as a function of time when the signal Hamiltonian is given by $\hat{H}=\prod_{i=1}^{5}\hat{Z}_{i}$, in the presence of local dephasing noise. In this case, by exploiting the repetition code defined in Eqs. \eqref{localcodewords1} and \eqref{localcodewords2}, AutoQEC order up to $c=2$ can be achieved. For the numerics, we consider different AutoQEC orders $c$, with $R=100$. For $c=1$, we consider $\keto{\Phi_{q}}=(\keto{\mu_{0}}+\keto{\mu_{1}})/\sqrt{2}~\forall~ 1 \leq q \leq q_{\mathrm{max}}$. Other conventions are identical with those of Fig.~\ref{fig:correlatedepol}.} 
    \label{fig:zzzzzz}
\end{figure}

In the second numerical analysis, we employ the signal $\hat{H}=\prod_{i=1}^{N}\hat{Z}_{i}$, in the presence of \textit{local} dephasing noise where $[\mathbf{C}]_{ij}=\delta_{ij}$. This scenario satisfies conditions in Theorem \ref{theorem1} up to AutoQEC order $1 \leq c \leq \lfloor{\frac{N-1}{2}}\rfloor$ with the code words
\begin{align}
    &\keto{\mu_{0}}=(\keto{+}^{\otimes N}+\keto{-}^{\otimes N})/\sqrt{2},\label{localcodewords1}\\
    &\keto{\mu_{1}}=(\keto{+}^{\otimes N}-\keto{-}^{\otimes N})/\sqrt{2}.\label{localcodewords2}
\end{align}
We considered a $5$-qubit system with $w=1$, $\kappa=0.1$, and $R=100$.
In Fig.~\ref{fig:zzzzzz}, we show that HS can be restored by AutoQEC. 
In particular, larger $c$ can more efficiently restore HS with the same $R$. 
A more detailed analysis of the numerics is found in the SM Sec.~S4.

\section{Discussion} We have established a sufficient condition ensuring that AutoQEC can efficiently restore HS. Notably, our code does not require noiseless ancilla and furthermore enables higher-order AutoQEC, both of which are crucial for its practical implementation. {Notably, our code does not require noiseless ancillas and furthermore enables higher-order AutoQEC, both of which can constitute important practical advantages, especially in settings where noiseless ancillas, large measurement–feedback overhead, or the implementation of sufficiently large recovery rates $R$ are difficult to realize.} AutoQEC has already been experimentally demonstrated in quantum computation \cite{aqecex-gertler2021protecting, aqecex-grimm2020stabilization, aqecex-lescanne2020exponential, aqecex-li2024autonomous}. 
In quantum metrology, a theoretical proposal for the implementation of AutoQEC has also been put forward \cite{con-reiter2017dissipative}, and we anticipate future experiments to demonstrate AutoQEC-based quantum estimation, thus overcoming limitations by noise in quantum metrology applications.  

We finally emphasize the major differences of our study from Ref. \cite{aqec-lebreuilly2021autonomousquantumerrorcorrection} which investigate AutoQEC in the context of quantum computation. Ref. \cite{aqec-lebreuilly2021autonomousquantumerrorcorrection} demonstrates that AutoQEC can efficiently suppress noise during computation when the Knill-Laflamme condition is satisfied, by proposing methods for constructing unitary operations corresponding to logical gates. In contrast, we propose an ancilla-free AutoQEC code for quantum metrology, tailored to scenarios in which the unitary operation responsible for imprinting the signal is given. This highlights the essential distinction between AutoQEC in computation and in metrology, thereby consolidating the unique contribution of our work.

\section{Methods}
\textbf{Proof of Theorem \ref{theorem1}}.

\noindent \textit{Step 1.--- Existence of code:} First, we suggest a code that satisfies the Knill-Laflamme condition for error set $\mathcal{E}^{[\sim c]}$. Let us denote the vectors satisfying the linear equation of (T2) as $\vb*{p}_{0}:=([\vb*{p}_{0}]_{1},[\vb*{p}_{0}]_{2},\cdots,[\vb*{p}_{0}]_{N_{0}})^{\mathrm{T}}$, $\vb*{p}_{1}:=([\vb*{p}_{1}]_{1},[\vb*{p}_{1}]_{2},\cdots,[\vb*{p}_{1}]_{N_{1}})^{\mathrm{T}}$. While multiple $\vb*{p}_{0} \in \mathbb{R}^{N_{i}}_{\geq 0}$, $\vb*{p}_{1} \in \mathbb{R}^{N_{j}}_{\geq 0}$ may exist satisfying the linear equation $\mathbf{A}_{0}\cdot \vb*{p}_{0}=\mathbf{A}_{1}\cdot \vb*{p}_{1}$, our proof does not depend on the specific choice of $(\vb*{p}_{0},\vb*{p}_{1})$. Therefore, we consider an arbitrary $(\vb*{p}_{0},\vb*{p}_{1})$ that satisfies the homogeneous equation in (T2). Based on $(\vb*{p}_{0},\vb*{p}_{1})$, we consider the following two-dimensional code $\mathcal{C}$, with the codewords specified as follows:
\begin{align}
    \ket{\mu_{0}}:=\sum_{k=1}^{N_{0}}{\sqrt{[\vb*{p}_{0}]_{k}}}\keto{h^{(k)}_{0}},~~\ket{\mu_{1}}:=\sum_{k=1}^{N_{1}}{\sqrt{[\vb*{p}_{1}]_{k}}}\keto{h^{(k)}_{1}}. \label{codewords}
\end{align}
Using Eqs. \eqref{formethod1}-\eqref{formethod3}, one can easily verify that the existence of $(\vb*{p}_{0},\vb*{p}_{1})$ implies
$\brao{\mu_{0}}\hat{E}^{\dagger}_{a}\hat{E}_{b}\keto{\mu_{0}}=\brao{\mu_{1}}\hat{E}^{\dagger}_{a}\hat{E}_{b}\keto{\mu_{1}}~\forall~ \hat{E}_{a},\hat{E}_{b} \in \mathcal{E}^{[\sim c]}$. In addition, the first condition $[\hat{H},\hat{L}_{\text{n},i}]=0~ \forall~ i$ ensures that $\hat{E}^{\dagger}_{a}\hat{E}_{b}~\forall~ \hat{E}_{a},\hat{E}_{b} \in \mathcal{E}^{[\sim c]}$ and $\hat{H}$ commute, resulting $\brao{\mu_{0}} \hat{E}^{\dagger}_{a}\hat{E}_{b} \keto{\mu_{1}}=\brao{\mu_{1}} \hat{E}^{\dagger}_{a}\hat{E}_{b} \keto{\mu_{0}}=0~\forall~ \hat{E}_{a},\hat{E}_{b} \in \mathcal{E}^{[\sim c]}$. This can be easily proven using the following equation:
\begin{align}
    \begin{split}
    &\brao{\mu_{0}} \hat{H}\hat{E}^{\dagger}_{a}\hat{E}_{b} \keto{\mu_{1}}=h_{0}\brao{\mu_{0}} \hat{E}^{\dagger}_{a}\hat{E}_{b} \keto{\mu_{1}}= \brao{\mu_{0}}\hat{E}^{\dagger}_{a}\hat{E}_{b}  \hat{H} \keto{\mu_{1}} \\
    &= h_{1}\brao{\mu_{0}} \hat{E}^{\dagger}_{a}\hat{E}_{b} \keto{\mu_{1}}~\forall~ \hat{E}_{a},\hat{E}_{b} \in \mathcal{E}^{[\sim c]}. 
    \end{split}\label{eqkloffdiag}
\end{align}
Here, since $h_{0} \neq h_{1}$, Eq. \eqref{eqkloffdiag} is valid if and only if $\brao{\mu_{0}} \hat{E}^{\dagger}_{a}\hat{E}_{b} \keto{\mu_{1}}=0~\forall~ \hat{E}_{a},\hat{E}_{b} \in \mathcal{E}^{[\sim c]}$.
As a consequence, we conclude that a code exists that satisfies the Knill-Laflamme condition for the error set $\mathcal{E}^{[\sim c]}$ if the (T1) and (T2) in Theorem. \ref{theorem1} are satisfied. 
$ $\newline
\textit{Step 2.--- Expression of Hamiltonian:}
For the second step of the proof, we note that $\keto{\mu_{0}}$ and $\keto{\mu_{1}}$ are eigenvectors of $\hat{H}$ with eigenvalues $h_{0}$ and $h_{1}$ respectively. Furthermore, according to the definition of the eigenbasis of the correctable error spaces, the eigenbasis $\keto{\mu^{[n]}_{i,i_{n}}}$ (where $i \in \{0,1\}$ and $1 \leq n \leq c$), can be expressed as a linear combination of $\{\hat{E}_{a}\keto{\mu_{i}}\}_{a}$ where $\hat{E}_{a} \in \mathcal{E}^{[\sim c]}$. Combining this with $[\hat{H},\hat{L}_{\text{n},i}]=0,~ \forall~ i$, it follows that $\hat{H}\keto{\mu^{[n]}_{i,i_{n}}}=h_{i}\keto{\mu^{[n]}_{i,i_{n}}}$ for $i \in \{0,1\}$. These relations imply the following key orthogonality conditions:
\begin{align}
    &\langle \mu^{[n]}_{i,i_{n}} \vert \hat{H} \vert \mu^{[m]}_{j,j_{m}} \rangle = h_{i}\delta_{i,j}\delta_{n,m}\delta_{i_{n},j_{m}},\\
    &\langle \phi_{q} \vert \hat{H} \vert \mu^{[m]}_{j,j_{m}} \rangle = \langle \phi_{q} \vert h_{j} \vert \mu^{[m]}_{j,j_{m}} \rangle = h_{j}\langle \phi_{q}    \vert \mu^{[m]}_{j,j_{m}} \rangle=0.
\end{align}
As a consequence, the Hamiltonian $\hat{H}$ can be expressed as
\begin{align}
    \begin{split}
    &\hat{H}=\sum_{j=0}^{1}\sum_{n=0}^{c}\sum_{i_{n}=1}^{p_{n}} h_{j} \vert \mu^{[n]}_{j,i_{n}} \rangle \langle  \mu^{[n]}_{j,i_{n}} \vert + \hat{\Pi}_{\mathrm{res}}\hat{H}\hat{\Pi}_{\mathrm{res}} \\
    &= \sum_{j,k=0}^{1}\sum_{n=0}^{c}\sum_{i_{n}=1}^{p_{n}} \brao{\mu_{j}} \hat{H}_{0} \keto{\mu_{k}} \vert \mu^{[n]}_{j,i_{n}} \rangle \langle  \mu^{[n]}_{k,i_{n}} \vert + \hat{\Pi}_{\mathrm{res}}\hat{H}\hat{\Pi}_{\mathrm{res}},
    \end{split}\label{method:Hform}
\end{align}
where $\hat{\Pi}_{\mathrm{res}}:= \sum_{q=1}^{q_{\mathrm{max}}}\dyado{\phi_{q}}$ is the projector of the residual space, and $\hat{H}_{0}:=\sum_{j=0}^{1}h_{j}\keto{\mu_{j}}\brao{\mu_{j}}$. 

To investigate whether there exists an engineered dissipation that achieves AutoQEC up to order $c$ for quantum metrology when (T1) and (T2) are satisfied, let us revisit Lemma \ref{performAQEC} in the main text which is the preliminary result of Ref. \cite{aqec-lebreuilly2021autonomousquantumerrorcorrection}. In Lemma \ref{performAQEC}, they state that if the Knill-Laflamme condition is satisfied for the error set $\mathcal{E}^{[\sim c]}$, the AutoQEC up to order $c$ can be achieved by applying the following engineered dissipation operation and the Hamiltonian
\begin{align}
    &\tilde{\mathcal{L}}_{\mathrm{E}} = \sum_{n=1}^{c}\sum_{i_{n}=1}^{p_{n}}D[\hat{L}^{[n]}_{\mathrm{E},i_{n}}]+ \sum_{q=1}^{q_{\mathrm{max}}}D[\hat{L}^{[\mathrm{res}]}_{\mathrm{E},q}], \label{method:engineereddissi}\\
    &\hat{H}=\sum_{j,k=1}^{d_{\mathcal{C}}}\sum_{n=0}^{c}\sum_{i_{n}=1}^{p_{n}} \brao{ \mu_{j} } \hat{H}_{0} \keto{ \mu_{k} } \keto{ \mu^{[n]}_{j,i_{n}} } \brao{  \mu^{[n]}_{k,i_{n}} }, \label{method:AQECHamil} 
\end{align}
where $\hat{L}^{[n]}_{\mathrm{E},i_{n}}=\sum_{j=0}^{d_{\mathcal{C}}-1}\vert \mu_{j} \rangle \langle \mu^{[n]}_{j,i_{n}} \vert ~ \mathrm{  for }~ 1\leq n \leq c ~\mathrm{ and }~ 1\leq i_{n} \leq p_{n}$, and $\hat{L}^{[\mathrm{res}]}_{\mathrm{E},q}= \vert \Phi_{q} \rangle \langle \phi_{q} \vert ~\mathrm{  for } ~1\leq q \leq q_{\mathrm{max}}$. Here, $\keto{\Phi_{q}} $ can be chosen among any normalized states in $\mathcal{C}$.
Notably, the only difference between the Hamiltonian designed for AutoQEC in computation (which is Eq. \eqref{method:AQECHamil}) and the given signal Hamiltonian (which is Eq. \eqref{method:Hform}) is the term $\hat{\Pi}_{\mathrm{res}}\hat{H}\hat{\Pi}_{\mathrm{res}}$. This term does not affect the performance of AutoQEC since it operates exclusively within the residual space $\mathcal{R}$. As a consequence, by applying the engineered dissipation defined in Eq. \eqref{method:engineereddissi}, AutoQEC up to order $c$ can be performed. Therefore, AutoQEC state $\hat{\rho}(t)$ satisfies
\begin{align}
    \left\lVert   \tilde{\mathcal{P}}_{\mathrm{E}}\left[\hat{\rho}(t)\right]-\hat{\rho}_{\mathrm{id}}(t) \right\rVert \leq M \left\lVert \hat{\rho}(0) \right\rVert \frac{\kappa t}{R^{c}} \label{method:aqeccondi}
\end{align}
for all $t,R \geq 0$, where $\hat{\rho}_{\mathrm{id}}(t)=\hat{U}_{0}(wt)\hat{\rho}_{0}\hat{U}^{\dagger}_{0}(wt)$ with $\hat{U}_{0}(wt)=e^{-iwt\hat{H}_{0}}$ and $\hat{\rho}_{0}$ is the initial quantum probe.

 $ $\newline
\textit{Step 3.--- QFI:} Until now, we have demonstrated that if the sufficient condition in Theorem \ref{theorem1} is satisfied, AutoQEC up to order $c$ can be performed. Lastly, we show that if AutoQEC up to order $c$ is achieved, then, the QFI of the AutoQEC state achieves HS.

We note that AutoQEC up to order $c$ is characterized by the operator norm of the difference between two states. Let us consider two states $\hat{\rho}$ and $\hat{\sigma}$ for further analysis. The difference of the two states can be expressed as
\begin{align}
    \hat{\rho}-\hat{\sigma}=\sum_{i}\lambda_{i}\dyado{\lambda_{i}},
\end{align}
where $\keto{\lambda_{i}}$ are the eigenvectors of $\hat{\rho}-\hat{\sigma}$ and the eigenvalues $\lambda_{i}$ are ordered such that $\abs{\lambda_{i}} \geq \abs{\lambda_{i+1}}$. The operator norm of $\hat{\rho}-\hat{\sigma}$ is then given by $\left\lVert \hat{\rho}-\hat{\sigma} \right\rVert = \abs{\lambda_{1}}:=x$. Since $x \geq \abs{\lambda_{i}} \geq \abs{\lambda_{i+1}}$, $\hat{\rho}-\hat{\sigma}$ can be expressed as
\begin{align}
    \hat{\rho}-\hat{\sigma}= \sum_{k=1}^{\infty}\hat{M}_{k}x^{k}.
\end{align}

As a consequence, the corresponding QFI difference can also be expressed as
\begin{align}
    F[\hat{\rho}]  - F[\hat{\sigma}]=\sum_{k=1}^{\infty}F_{k}x^{k}.
\end{align}
Here, we note that since $\sum_{k=1}^{\infty}F_{k}x^{k} \vert_{x=0}=0$,
\begin{align}
    \exists M_{0}>0,~\exists x_{0}~\forall~ 0\leq x<x_{0}: \abs{\sum_{k=1}^{\infty}F_{k}x^{k}} \leq M_{0}x. \label{method:bigomega}
\end{align}
Next, let us consider $\hat{\rho}=\tilde{\mathcal{P}}_{\mathrm{E}}\left[\hat{\rho}(t)\right]$ and $\hat{\sigma}=\hat{\rho}_{\mathrm{id}}(t)$. We again emphasize that Eq. \eqref{method:bigomega} holds for $x<x_{0}$ and according to Eq. \eqref{method:aqeccondi}, we have $x \leq  M \left\lVert \hat{\rho}(0) \right\rVert \frac{\kappa t}{R^{c}} \leq  M \left\lVert \hat{\rho}(0) \right\rVert \frac{\kappa T}{R^{c}}$. Therefore, for $R$ satisfying $M \left\lVert \hat{\rho}(0) \right\rVert \frac{\kappa T}{R^{c}} <x_{0}$, equivalently $\left( {M \left\lVert \hat{\rho}(0) \right\rVert \kappa T}/{x_{0}}\right)^{\frac{1}{c}} < R$, we have
\begin{align}
    \abs{F[\tilde{\mathcal{P}}_{\mathrm{E}}\left[\hat{\rho}(t)\right]]-F[\hat{\rho}_{\mathrm{id}}(t)]} \leq M_{0} M \left\lVert \hat{\rho}(0) \right\rVert  \frac{\kappa t}{R^{c}} \leq M_{0} M  \left\lVert \hat{\rho}(0) \right\rVert \frac{\kappa T}{R^{c}}. 
\end{align}
For a given arbitrarily small error $\epsilon >0$, by choosing $R$ such that $\left({M_{0}M \left\lVert \hat{\rho}(0) \right\rVert \kappa T}/{\epsilon}\right)^{\frac{1}{c}} = R$, we have 
\begin{align}
    \abs{F[\tilde{\mathcal{P}}_{\mathrm{E}}\left[\hat{\rho}(t)\right]]-F[\hat{\rho}_{\mathrm{id}}(t)]} \leq M_{0} M \left\lVert \hat{\rho}(0) \right\rVert  \frac{\kappa T}{R^{c}} = \epsilon. \label{method:pqfibound}
\end{align}
Next, we derive Eq. \eqref{eqtheorem} from Eq. \eqref{method:pqfibound}. In this derivation, we consider the scenario in which the QFI of the AutoQEC state is always less than or equal to that of the ideal case over the entire sensing duration, i.e., $F[\hat{\rho}(t)] \leq F[\hat{\rho}_{\mathrm{id}}(t)]$ for all $0 \leq t \leq T$. (We note that in cases where there exists some $t$ such that $F[\hat{\rho}(t)] > F[\hat{\rho}_{\mathrm{id}}(t)]$, the derivation remains straightforward.) In this case, since $\tilde{\mathcal{P}}_{\mathrm{E}}$ is a well-defined CPTP map, the data processing inequality of the QFI ensures the following inequality holds \cite{est-petz2007quantum}:
\begin{align}
    F[\tilde{\mathcal{P}}_{\mathrm{E}}\left[\hat{\rho}(t)\right]] \leq F[\hat{\rho}(t)] \leq F[\hat{\rho}_{\mathrm{id}}(t)]. \label{method:qfiinequality}
\end{align}
We then obtain the following equation which can be easily derived from Eqs. \eqref{method:pqfibound} and \eqref{method:qfiinequality}:
\begin{align}
    F[\hat{\rho}_{\mathrm{id}}(t)]-F[\hat{\rho}(t)] \leq \epsilon. \label{method:lastqfi}
\end{align}
To proceed, we consider the initial quantum probe state $\keto{\mu_{0}}+\keto{\mu_{1}}$, for which the QFI of the ideal state is given by $F[\hat{\rho}_{\mathrm{id}}(t)] = (h_{0}-h_{1})^{2}t^{2}$. Consequently, substituting this expression into Eq. \eqref{method:lastqfi} leads directly to Eq. \eqref{eqtheorem}.

\section{Acknowledgments}
We acknowledge useful discussions with Ivan Rojkov, Sisi Zhou and Kyungjoo Noh. 
This research was funded by National Research Foundation of Korea (RS-2022-NR068812), and Institute of Information \& Communications Technology Planning \& Evaluation (RS-2025-02263264).
H.K. is supported by the Education and Training Program of the Quantum Information Research Support Center (No.~2021M3H3A1036573), the IITP (RS-2025-25464252, RS-2024-00437191), the NRF (RS-2025-25464492, RS-2024-00442710) funded by the Ministry of Science and ICT (MSIT)  Korea.
U.R.F. was supported by the NRF under Grant No.~2020R1A2C2008103. 
L.J. acknowledges support from the ARO(W911NF-23-1-0077), ARO MURI (W911NF-21-1-0325), AFOSR MURI (FA9550-19-1-0399, FA9550-21-1-0209, FA9550-23-1-0338), DARPA (HR0011-24-9-0359, HR0011-24-9-0361), NSF (OMA-1936118, ERC-1941583, OMA-2137642, OSI-2326767, CCF-2312755), NTT Research, and the Packard Foundation (2020-71479).

\bibliography{Reference.bib}

\let\oldaddcontentsline\addcontentsline
\renewcommand{\addcontentsline}[3]{}
\let\addcontentsline\oldaddcontentsline
\onecolumngrid

\clearpage
\begin{center}
	\Large
	\textbf{Supplemental Material: Restoring Heisenberg scaling in time via autonomous quantum error correction}
\end{center}

\setcounter{equation}{0}
\setcounter{figure}{0}
\setcounter{table}{0}

\setcounter{page}{1}
\renewcommand{\thesection}{S\arabic{section}}
\renewcommand{\theequation}{S\arabic{equation}}
\renewcommand{\thefigure}{S\arabic{figure}}
\renewcommand{\thetable}{S\arabic{table}}

\addtocontents{toc}{\protect\setcounter{tocdepth}{0}}

\setcounter{section}{0}

\setcounter{section}{0}

\section{Proof of the Theorem 1}
In the main text, we express the signal Hamiltonian $\hat{H}$ in the diagonalized form as $\hat{H}=\sum_{i=0}^{d-1}h_{i}\left(\sum_{l=1}^{N_{i}}\dyado{h^{(l)}_{i}}\right)$,
where $\{h_{i}\}_{i=0}^{d-1}$ are the distinct eigenvalues, and $\{\keto{h^{(k)}_{i}}\}_{k=1}^{N_{i}}$ are eigenvectors with eigenvalue $h_{i}$. Based on the expression, we define the following operator set: 
\begin{align}
    \mathcal{K}^{[\sim c]}:=\{\hat{E}^{\dagger}_{a}\hat{E}_{b} \vert \hat{E}_{a},\hat{E}_{b} \in \mathcal{E}^{[\sim c]}~\forall~ a,b \}. \label{defofK}
\end{align}
According to the definition of $\mathcal{K}^{[\sim c]}$, the Knill-Laflamme condition for the error set $\mathcal{E}^{[\sim c]}$ can be expressed as 
\begin{align}
    \brao{\mu_{i}}\hat{K}_{k}\keto{\mu_{j}}=\sigma_{k}\delta_{ij}~\forall~ 1 \leq k \leq \abs{\mathcal{K}^{[\sim c]}},
\end{align}
where $\hat{K}_{k} \in \mathcal{K}^{[\sim c]}$ for all $k$, and $\sigma_{k}$ are constants associated with the error operators.
Next, using $\mathcal{K}$ and the eigenvectors of $\hat{H}$, we define $\abs{\mathcal{K}^{[\sim c]}} \times N_{i}$ matrix $\mathbf{A}^{[\sim c]}_{i}$, where $(k,l)$-elements are given by
\begin{align}
[\mathbf{A}^{[\sim c]}_{i}]_{kl} :=\brao{h^{(l)}_{i}} \hat{K}_{k} \keto{h^{(l)}_{i}}~\forall~ k,l. \label{defofA}
\end{align}
Based on $\mathbf{A}^{[\sim c]}_{i}$, we introduce the following Theorem:
\setcounter{theorem}{0}
\begin{theorem} \label{supple:theorem1}
    Suppose the following two conditions hold: (T1) $[\hat{H},\hat{L}_{\text{n},a}]=0~ \forall~ a$, and (T2) $\exists i,j \neq i$, along with two probability vectors $\vb*{p}_{i} \in \mathbb{R}^{N_{i}}_{\geq 0}$, $\vb*{p}_{j} \in \mathbb{R}^{N_{j}}_{\geq 0}$ such that $\mathbf{A}^{[\sim c]}_{i}\cdot \vb*{p}_{i}=\mathbf{A}^{[\sim c]}_{j}\cdot \vb*{p}_{j}$.
    
    Then, for any given time $T \geq 0$, and arbitrarily small error $\epsilon>0$, there exists an ancilla-free AutoQEC scheme with a finite $R$ such that
    \begin{align}
    \forall~ 0\leq t \leq T:~F[\hat{\rho}(t)] \geq (h_{i}-h_{j})^{2}t^{2} - \epsilon, \label{supple:eqtheorem}
    \end{align}
    where the relationship between $R$, $\kappa$, $T$, and $\epsilon$ is given by $\epsilon=O(\kappa T / R^{c})$. Here, $\hat{\rho}(t)$ is the AutoQEC state, and $F[\hat{\rho}]$ is the QFI of $\hat{\rho}$ with respect to $w$.
\end{theorem}
In the rest of the proof, for simplicity, let us assume that $p_{0}$ and $p_{1}$ satisfy (T2), without loss of generality.

\begin{proof}
 $ $\newline
\textit{Step 1.--- Existence of code:} First, we suggest a code that satisfies the Knill-Laflamme condition for error set $\mathcal{E}^{[\sim c]}$. Let us denote the vectors satisfying the linear equation of (T2) as $\vb*{p}_{0}:=([\vb*{p}_{0}]_{1},[\vb*{p}_{0}]_{2},\cdots,[\vb*{p}_{0}]_{N_{0}})^{\mathrm{T}}$, $\vb*{p}_{1}:=([\vb*{p}_{1}]_{1},[\vb*{p}_{1}]_{2},\cdots,[\vb*{p}_{1}]_{N_{1}})^{\mathrm{T}}$. While multiple $\vb*{p}_{0} \in \mathbb{R}^{N_{i}}_{\geq 0}$, $\vb*{p}_{1} \in \mathbb{R}^{N_{j}}_{\geq 0}$ may exist satisfying the linear equation $\mathbf{A}_{0}\cdot \vb*{p}_{0}=\mathbf{A}_{1}\cdot \vb*{p}_{1}$, our proof does not depend on the specific choice of $(\vb*{p}_{0},\vb*{p}_{1})$. Therefore, we consider an arbitrary $(\vb*{p}_{0},\vb*{p}_{1})$ that satisfies the homogeneous equation in (T2). Based on $(\vb*{p}_{0},\vb*{p}_{1})$, we consider the following two-dimensional code $\mathcal{C}$, with the codewords specified as follows:
\begin{align}
    \ket{\mu_{0}}:=\sum_{k=1}^{N_{0}}{\sqrt{[\vb*{p}_{0}]_{k}}}\keto{h^{(k)}_{0}},~~\ket{\mu_{1}}:=\sum_{k=1}^{N_{1}}{\sqrt{[\vb*{p}_{1}]_{k}}}\keto{h^{(k)}_{1}}. \label{codewords}
\end{align}
Using Eqs. \eqref{defofK}-\eqref{defofA}, one can easily verify that the existence of $(\vb*{p}_{0},\vb*{p}_{1})$ implies
$\brao{\mu_{0}}\hat{E}^{\dagger}_{a}\hat{E}_{b}\keto{\mu_{0}}=\brao{\mu_{1}}\hat{E}^{\dagger}_{a}\hat{E}_{b}\keto{\mu_{1}}~\forall~ \hat{E}_{a},\hat{E}_{b} \in \mathcal{E}^{[\sim c]}$. In addition, the first condition $[\hat{H},\hat{L}_{\text{n},i}]=0~ \forall~ i$ ensures that $\hat{E}^{\dagger}_{a}\hat{E}_{b}~\forall~ \hat{E}_{a},\hat{E}_{b} \in \mathcal{E}^{[\sim c]}$ and $\hat{H}$ commute, resulting $\brao{\mu_{0}} \hat{E}^{\dagger}_{a}\hat{E}_{b} \keto{\mu_{1}}=\brao{\mu_{1}} \hat{E}^{\dagger}_{a}\hat{E}_{b} \keto{\mu_{0}}=0~\forall~ \hat{E}_{a},\hat{E}_{b} \in \mathcal{E}^{[\sim c]}$. This can be easily proven using the following equation:
\begin{align}
    \brao{\mu_{0}} \hat{H}\hat{E}^{\dagger}_{a}\hat{E}_{b} \keto{\mu_{1}}=h_{0}\brao{\mu_{0}} \hat{E}^{\dagger}_{a}\hat{E}_{b} \keto{\mu_{1}}= \brao{\mu_{0}}\hat{E}^{\dagger}_{a}\hat{E}_{b}  \hat{H} \keto{\mu_{1}} = h_{1}\brao{\mu_{0}} \hat{E}^{\dagger}_{a}\hat{E}_{b} \keto{\mu_{1}}~\forall~ \hat{E}_{a},\hat{E}_{b} \in \mathcal{E}^{[\sim c]}. \label{kloffdiag}
\end{align}
Here, since $h_{0} \neq h_{1}$, Eq. \eqref{kloffdiag} is valid if and only if $\brao{\mu_{0}} \hat{E}^{\dagger}_{a}\hat{E}_{b} \keto{\mu_{1}}=0~\forall~ \hat{E}_{a},\hat{E}_{b} \in \mathcal{E}^{[\sim c]}$.
As a consequence, we conclude that a code exists that satisfies the Knill-Laflamme condition for the error set $\mathcal{E}^{[\sim c]}$ if the (T1) and (T2) in Theorem. \ref{supple:theorem1} are satisfied. 
$ $\newline
\textit{Step 2.--- Expression of Hamiltonian:}
For the second step of the proof, we note that $\keto{\mu_{0}}$ and $\keto{\mu_{1}}$ are eigenvectors of $\hat{H}$ with eigenvalues $h_{0}$ and $h_{1}$ respectively. Furthermore, according to the definition of the eigenbasis of the correctable error spaces, the eigenbasis $\keto{\mu^{[n]}_{i,i_{n}}}$ (where $i \in \{0,1\}$ and $1 \leq n \leq c$), can be expressed as a linear combination of $\{\hat{E}_{a}\keto{\mu_{i}}\}_{a}$ where $\hat{E}_{a} \in \mathcal{E}^{[\sim c]}$. Combining this with $[\hat{H},\hat{L}_{\text{n},i}]=0,~ \forall~ i$, it follows that $\hat{H}\keto{\mu^{[n]}_{i,i_{n}}}=h_{i}\keto{\mu^{[n]}_{i,i_{n}}}$ for $i \in \{0,1\}$. These relations imply the following key orthogonality conditions:
\begin{align}
    \langle \mu^{[n]}_{i,i_{n}} \vert \hat{H} \vert \mu^{[m]}_{j,j_{m}} \rangle = h_{i}\delta_{i,j}\delta_{n,m}\delta_{i_{n},j_{m}},\quad    \langle \phi_{q} \vert \hat{H} \vert \mu^{[m]}_{j,j_{m}} \rangle = \langle \phi_{q} \vert h_{j} \vert \mu^{[m]}_{j,j_{m}} \rangle = h_{j}\langle \phi_{q}    \vert \mu^{[m]}_{j,j_{m}} \rangle=0.
\end{align}
As a consequence, the Hamiltonian $\hat{H}$ can be expressed as
\begin{align}
    \hat{H}=\sum_{j=0}^{1}\sum_{n=0}^{c}\sum_{i_{n}=1}^{p_{n}} h_{j} \vert \mu^{[n]}_{j,i_{n}} \rangle \langle  \mu^{[n]}_{j,i_{n}} \vert + \hat{\Pi}_{\mathrm{res}}\hat{H}\hat{\Pi}_{\mathrm{res}} = \sum_{j,k=0}^{1}\sum_{n=0}^{c}\sum_{i_{n}=1}^{p_{n}} \brao{\mu_{j}} \hat{H}_{0} \keto{\mu_{k}} \vert \mu^{[n]}_{j,i_{n}} \rangle \langle  \mu^{[n]}_{k,i_{n}} \vert + \hat{\Pi}_{\mathrm{res}}\hat{H}\hat{\Pi}_{\mathrm{res}}, \label{supple:Hform}
\end{align}
where $\hat{\Pi}_{\mathrm{res}}:= \sum_{q=1}^{q_{\mathrm{max}}}\dyado{\phi_{q}}$ is the projector of the residual space, and $\hat{H}_{0}:=\sum_{j=0}^{1}h_{j}\keto{\mu_{j}}\brao{\mu_{j}}$. 

To investigate whether there exists an engineered dissipation that achieves AutoQEC up to order $c$ for quantum metrology when (T1) and (T2) are satisfied, let us revisit Lemma 1 in the main text which is the preliminary result of Ref. \cite{aqec-lebreuilly2021autonomousquantumerrorcorrection}. In Lemma 1, they state that if the Knill-Laflamme condition is satisfied for the error set $\mathcal{E}^{[\sim c]}$, the AutoQEC up to order $c$ can be achieved by applying the following engineered dissipation operation and the Hamiltonian
\begin{align}
    &\tilde{\mathcal{L}}_{\mathrm{E}} = \sum_{n=1}^{c}\sum_{i_{n}=1}^{p_{n}}D[\hat{L}^{[n]}_{\mathrm{E},i_{n}}]+ \sum_{q=1}^{q_{\mathrm{max}}}D[\hat{L}^{[\mathrm{res}]}_{\mathrm{E},q}], \label{supple:engineereddissi}\\
    &\hat{H}=\sum_{j,k=0}^{d_{\mathcal{C}}-1}\sum_{n=0}^{c}\sum_{i_{n}=1}^{p_{n}} \brao{ \mu_{j} } \hat{H}_{0} \keto{ \mu_{k} } \keto{ \mu^{[n]}_{j,i_{n}} } \brao{  \mu^{[n]}_{k,i_{n}} }, \label{AQECHamil} 
\end{align}
where $\hat{L}^{[n]}_{\mathrm{E},i_{n}}=\sum_{j=0}^{d_{\mathcal{C}}-1}\vert \mu_{j} \rangle \langle \mu^{[n]}_{j,i_{n}} \vert ~ \mathrm{  for }~ 1\leq n \leq c ~\mathrm{ and }~ 1\leq i_{n} \leq p_{n}$, and $\hat{L}^{[\mathrm{res}]}_{\mathrm{E},q}= \vert \Phi_{q} \rangle \langle \phi_{q} \vert ~\mathrm{  for } ~1\leq q \leq q_{\mathrm{max}}$. Here, $\keto{\Phi_{q}} $ can be chosen among any normalized states in $\mathcal{C}$.
Notably, the only difference between the Hamiltonian designed for AutoQEC in computation (which is Eq. \eqref{AQECHamil}) and the given signal Hamiltonian (which is Eq. \eqref{supple:Hform}) is the term $\hat{\Pi}_{\mathrm{res}}\hat{H}\hat{\Pi}_{\mathrm{res}}$. This term does not affect the performance of AutoQEC since it operates exclusively within the residual space $\mathcal{R}$. As a consequence, by applying the engineered dissipation defined in Eq. \eqref{supple:engineereddissi}, AutoQEC up to order $c$ can be performed. Therefore, AutoQEC state $\hat{\rho}(t)$ satisfies
\begin{align}
    \left\lVert   \tilde{\mathcal{P}}_{\mathrm{E}}\left[\hat{\rho}(t)\right]-\hat{\rho}_{\mathrm{id}}(t) \right\rVert \leq M \left\lVert \hat{\rho}(0) \right\rVert \frac{\kappa t}{R^{c}} \label{supple:aqeccondi}
\end{align}
for all $t,R \geq 0$, where $\hat{\rho}_{\mathrm{id}}(t)=\hat{U}_{0}(wt)\hat{\rho}_{0}\hat{U}^{\dagger}_{0}(wt)$ with $\hat{U}_{0}(wt)=e^{-iwt\hat{H}_{0}}$ and $\hat{\rho}_{0}$ is the initial quantum probe.

 $ $\newline
\textit{Step 3.--- QFI:} Until now, we have demonstrated that if the sufficient condition in Theorem \ref{supple:theorem1} is satisfied, AutoQEC up to order $c$ can be performed. Lastly, we show that if AutoQEC up to order $c$ is achieved, then, the QFI of the AutoQEC state achieves HS.

We note that AutoQEC up to order $c$ is characterized by the operator norm of the difference between two states. Let us consider two states $\hat{\rho}$ and $\hat{\sigma}$ for further analysis. The difference of the two states can be expressed as
\begin{align}
    \hat{\rho}-\hat{\sigma}=\sum_{i}\lambda_{i}\dyado{\lambda_{i}},
\end{align}
where $\keto{\lambda_{i}}$ are the eigenvectors of $\hat{\rho}-\hat{\sigma}$ and the eigenvalues $\lambda_{i}$ are ordered such that $\abs{\lambda_{i}} \geq \abs{\lambda_{i+1}}$. The operator norm of $\hat{\rho}-\hat{\sigma}$ is then given by $\left\lVert \hat{\rho}-\hat{\sigma} \right\rVert = \abs{\lambda_{1}}:=x$. Since $x \geq \abs{\lambda_{i}} \geq \abs{\lambda_{i+1}}$, $\hat{\rho}-\hat{\sigma}$ can be expressed as
\begin{align}
    \hat{\rho}-\hat{\sigma}= \sum_{k=1}^{\infty}\hat{M}_{k}x^{k}.
\end{align}

As a consequence, the corresponding QFI difference can also be expressed as
\begin{align}
    F[\hat{\rho}]  - F[\hat{\sigma}]=\sum_{k=1}^{\infty}F_{k}x^{k}.
\end{align}
Here, we note that since $\sum_{k=1}^{\infty}F_{k}x^{k} \vert_{x=0}=0$,
\begin{align}
    \exists M_{0}>0,~\exists x_{0}~\forall~ 0\leq x<x_{0}: \abs{\sum_{k=1}^{\infty}F_{k}x^{k}} \leq M_{0}x. \label{bigomega}
\end{align}
Next, let us consider $\hat{\rho}=\tilde{\mathcal{P}}_{\mathrm{E}}\left[\hat{\rho}(t)\right]$ and $\hat{\sigma}=\hat{\rho}_{\mathrm{id}}(t)$. We again emphasize that Eq. \eqref{bigomega} holds for $x<x_{0}$ and according to Eq. \eqref{supple:aqeccondi}, we have $x \leq  M \left\lVert \hat{\rho}(0) \right\rVert \frac{\kappa t}{R^{c}} \leq  M \left\lVert \hat{\rho}(0) \right\rVert \frac{\kappa T}{R^{c}}$. Therefore, for $R$ satisfying $M \left\lVert \hat{\rho}(0) \right\rVert \frac{\kappa T}{R^{c}} <x_{0}$, equivalently $\left( {M \left\lVert \hat{\rho}(0) \right\rVert \kappa T}/{x_{0}}\right)^{\frac{1}{c}} < R$, we have
\begin{align}
    \abs{F[\tilde{\mathcal{P}}_{\mathrm{E}}\left[\hat{\rho}(t)\right]]-F[\hat{\rho}_{\mathrm{id}}(t)]} \leq M_{0} M \left\lVert \hat{\rho}(0) \right\rVert  \frac{\kappa t}{R^{c}} \leq M_{0} M  \left\lVert \hat{\rho}(0) \right\rVert \frac{\kappa T}{R^{c}}. 
\end{align}
For a given arbitrarily small error $\epsilon >0$, by choosing $R$ such that $\left({M_{0}M \left\lVert \hat{\rho}(0) \right\rVert \kappa T}/{\epsilon}\right)^{\frac{1}{c}} = R$, we have 
\begin{align}
    \abs{F[\tilde{\mathcal{P}}_{\mathrm{E}}\left[\hat{\rho}(t)\right]]-F[\hat{\rho}_{\mathrm{id}}(t)]} \leq M_{0} M \left\lVert \hat{\rho}(0) \right\rVert  \frac{\kappa T}{R^{c}} = \epsilon. \label{pqfibound}
\end{align}
Next, we derive Eq. \eqref{supple:eqtheorem} from Eq. \eqref{pqfibound}. In this derivation, we consider the scenario in which the QFI of the AutoQEC state is always less than or equal to that of the ideal case over the entire sensing duration, i.e., $F[\hat{\rho}(t)] \leq F[\hat{\rho}_{\mathrm{id}}(t)]$ for all $0 \leq t \leq T$. (We note that in cases where there exists some $t$ such that $F[\hat{\rho}(t)] > F[\hat{\rho}_{\mathrm{id}}(t)]$, the derivation remains straightforward.) In this case, since $\tilde{\mathcal{P}}$ is a well-defined CPTP map, the data processing inequality of the QFI ensures the following inequality holds \cite{est-petz2007quantum}:
\begin{align}
    F[\tilde{\mathcal{P}}_{\mathrm{E}}\left[\hat{\rho}(t)\right]] \leq F[\hat{\rho}(t)] \leq F[\hat{\rho}_{\mathrm{id}}(t)]. \label{qfiinequality}
\end{align}
We then obtain the following equation which can be easily derived from Eqs. \eqref{pqfibound} and \eqref{qfiinequality}:
\begin{align}
    F[\hat{\rho}_{\mathrm{id}}(t)]-F[\hat{\rho}(t)] \leq \epsilon. \label{lastqfi}
\end{align}
To proceed, we consider the initial quantum probe state $\keto{\mu_{0}}+\keto{\mu_{1}}$, for which the QFI of the ideal state is given by $F[\hat{\rho}_{\mathrm{id}}(t)] = (h_{0}-h_{1})^{2}t^{2}$. Consequently, substituting this expression into Eq. \eqref{lastqfi} leads directly to Eq. \eqref{supple:eqtheorem}. 

Lastly, we note that the term $(h_{0}-h_{1})^{2}t^{2}$ does not depend on $\vb*{p}$, however, the constants $M_{0}$ and $x_{0}$ may depend on $\vb*{p}$, which is beyond our scope.
\end{proof}
We note that our sufficient condition is stricter compared to \textit{Hamiltonian-not-in-Lindblad-span} (HNLS) condition \cite{qec-zhou2018achieving}. Let us assume that HNLS condition is not satisfied, i.e., 
\begin{align}
    \hat{H} \in \mathrm{span}\{\mathcal{K}^{[\sim 1]}\}, \label{HNLSnotsati}
\end{align}
while our sufficient condition is satisfied. In this case, since the sufficient condition is satisfied, a code $\mathcal{C}$ with code words $\ket{\mu_{0,1}}$, satisfies the Knill-Laflamme condition as verified in the \textit{proof} of \textit{Step 1} of Theorem \ref{supple:theorem1}. At the same time, since we have assumed that HNLS condition is violated, the following equation should be satisfied: $\hat{\Pi}_{\mathcal{C}}\hat{H}\hat{\Pi}_{\mathcal{C}}=c\hat{\Pi}_{\mathcal{C}}$ where $c$ is constant and $\hat{\Pi}_{\mathcal{C}}= \sum_{i=0}^{1}\dyado{\mu_{i}}$ because of the satisfaction of the Knill-Laflamme condition and according to Eq. \eqref{HNLSnotsati}. However, since $\keto{\mu_{i}}$ are eigenvectors of $\hat{H}$ with different eigenvalues, this equation cannot be satisfied, leading to a contradiction. Consequently, we conclude that if HNLS is not satisfied, our sufficient condition also cannot be satisfied.

Finally, we further introduce the following \textit{corollary}, which pertains to the second numerical result presented in the main text. While the conditions in Corollary \ref{corollary2} are stricter than those in Theorem \ref{supple:theorem1}, they require less computational effort to verify their satisfaction, for example, the sensing scenario of the second numerical result in the main text.
\begin{corollary}\label{corollary2}
    AutoQEC can restore HS if all the following conditions are satisfied: (C1) $[\hat{H},\hat{L}_{\text{n},i}]=0,~ \forall~ i$, (C2) $\mathrm{Tr}[\hat{H}\hat{K}]=0$,  $\forall~ \hat{K} \in \mathcal{K}^{[\sim c]}$, (C3) $\hat{H}$ has only two distinct, non-zero eigenvalues. 
\end{corollary}
\textit{proof sketch.---} Let us denote the two non-zero eigenvalues as $h_{0}$ and $h_{1}$. If (C2) is satisfied, the signal Hamiltonian is a traceless Hermitian operator because $\hat{I} \in \mathcal{K}^{[\sim c]}$. Therefore, $\hat{H}$ can be expressed as
\begin{align}
    \hat{H}=\sum_{i=1}^{d}h_{0}\keto{h^{(k)}_{0}}+\sum_{i=1}^{d}h_{1}\keto{h^{(k)}_{1}}
\end{align}
where $h_{0}+h_{1}=0$.
We then consider the following code:
\begin{align}
    \keto{\mu_{0}}:=\frac{1}{\sqrt{d}}\sum_{k=1}^{d}\keto{h^{(k)}_{0}},~~ \keto{\mu_{1}}:=\frac{1}{\sqrt{d}}\sum_{k=1}^{d}\keto{h^{(k)}_{1}}. \label{codewords2}
\end{align}
The rest parts of the proof are the same as the proof of Theorem \ref{supple:theorem1}. \qed

\section{Linear Programming Approach to Detecting Condition (T2)} \label{Appen A}
In this section, we demonstrate that the solution to (T2) can be obtained using the \textit{linear programming} method. Moreover, if no solution exists, the infeasibility can also be detected through linear programming.

Linear programming (LP) is a mathematical optimization technique used to find the best possible outcome (such as maximum profit or minimum cost) in a mathematical model whose requirements are represented by linear relationships. 

We introduce the standard form of LP:
\begin{align}
    \mathrm{Maximize}:~f(\vb*{x})=\vb*{c}^{\mathrm{T}}\vb*{x},~~~\mathrm{subject~ to}: \mathbf{A}\vb*{x} \leq \vb*{b}, ~\vb*{x} \geq 0,
\end{align}
where $f(\vb*{x})$ is an objective function to be optimized, $\vb*{c}$ is a vector of coefficients in the objective function, $\mathbf{A}$ is a matrix of coefficients in the constraints, $\vb*{x}$ is the vector of decision variables, and $\vb*{b}$ is the vector of constants in the constraints.

To apply LP to solve the linear equation in (T2), we first show that the linear equation can be expressed as a homogeneous equation. Let us define $\abs{\mathcal{K}^{[\sim c]}} \times (N_{i}+N_{j})$ matrix $\mathbf{A}^{[\sim c]}_{ij}$, its $(k,l)$-elements are defined as
\begin{align}
\begin{cases}
  [\mathbf{A}^{[\sim c]}_{ij}]_{kl} :=\brao{h^{(l)}_{i}} \hat{K}_{k} \keto{h^{(l)}_{i}}, & 1 \leq l \leq N_{i}, \\
  [\mathbf{A}^{[\sim c]}_{ij}]_{k(l+N_{i})} :=-\brao{h^{(l)}_{j}} \hat{K}_{k} \keto{h^{(l)}_{j}}, & 1 \leq l \leq N_{j}. 
\end{cases}\label{defofA2}
\end{align}
Based on $\mathbf{A}^{[\sim c]}_{ij}$, (T2) can be re-expressed as $\exists i,j\neq i$, along with two probability vectors $\vb*{p}_{i} \in \mathbb{R}^{N_{i}}_{\geq 0}$, $\vb*{p}_{j} \in \mathbb{R}^{N_{j}}_{\geq 0}$ such that $\mathbf{A}^{[\sim c]}_{ij}\cdot \vb*{p}=0$ where $\vb*{p}:=(\vb*{p}_{i},\vb*{p}_{j})$. We then apply LP to obtain $\vb*{p}$ such that $\mathbf{A}^{[\sim c]}_{ij}\cdot \vb*{p}=0$, we consider following optimization problem:
\begin{align}
    \mathrm{Maximize}:~f(\vb*{p})=\vb*{0}^{\mathrm{T}}\vb*{p},~~~\mathrm{subject~ to}: ~~\Re\big(\mathbf{A}^{[\sim c]}_{ij}\big) \vb*{p}=0,~~\Im\big(\mathbf{A}^{[\sim c]}_{ij}\big)\vb*{p}=0,~~ \mathbf{B}_{i}\vb*{p} = \vb*{b}_{i},~~\mathbf{B}_{j}\vb*{p} = \vb*{b}_{j}, ~~\vb*{p} \geq 0, \label{LPforour}
\end{align}
where the relevant terms are defined as follows: $\mathbf{B}_{i}$ and 
$\mathbf{B}_{j}$ which are $\abs{\mathcal{K}^{[\sim c]}} \times (N_{i}+N_{j})$ matrices are defined as 
\begin{align}
\mathbf{B}_{i}=
\begin{cases}
  [\mathbf{B}_{i}]_{1l} :=1, & 1 \leq l \leq N_{i}, \\
  [\mathbf{B}_{i}]_{kl} :=0, & k\neq 1,~ (N_{i}+1) \leq l \leq (N_{i}+ N_{j}). 
\end{cases},~~
\mathbf{B}_{j}=
\begin{cases}
  [\mathbf{B}_{j}]_{1l} :=1, &  (N_{i}+1) \leq l \leq (N_{i}+ N_{j}), \\
  [\mathbf{B}_{j}]_{kl} :=0, & k\neq 1,~ 1 \leq l \leq N_{i}. 
\end{cases}.
\end{align}
$\vb*{b}_{i}$ is a vector defined with its components $[\vb*{b}_{i}]_{1}=1$ and $[\vb*{b}_{i}]_{k \neq 1}=0$, and $\vb*{b}_{j}$ is a vector defined with its components $[\vb*{b}_{j}]_{N_{i}+1}=1$ and $[\vb*{b}_{j}]_{k \neq (N_{i}+1)}=0$. According to these definitions, the constraints $\mathbf{B}_{i}\cdot\vb*{p} = \vb*{b}_{i},~\mathbf{B}_{j}\cdot\vb*{p} = \vb*{b}_{j}, ~\vb*{p} \geq 0,~~$ ensure that $\vb*{p}_{i}$ and $\vb*{p}_{j}$ are probability vectors satisfying
\begin{align}
    &\sum_{k=1}^{N_{i}}[\vb*{p}_{i}]_{k}=1,~~[\vb*{p}_{i}]_{k} \geq 0~\forall~ k,\\
    &\sum_{k=1}^{N_{j}}[\vb*{p}_{j}]_{k}=1,~~[\vb*{p}_{j}]_{k} \geq 0~\forall~ k.
\end{align}
Lastly, we note that when searching the existence of $h_{i}$ and $h_{j}$ satisfying (T2), it is more efficient to begin with the pairs ordered in descending magnitude of $\abs{h_{i}-h_{j}}$. This preference arises from our analysis in the \textit{proof} of Theorem \ref{supple:theorem1}, where we established that the QFI of the AutoQEC state is $(h_{i}-h_{j})^{2}t^{2}$.

The pseudo-code for the LP problem is given in Algorithm \ref{algo1}.
\begin{algorithm}[ht]\label{algo1}
\DontPrintSemicolon
\caption{Detecting nontrivial solutions in descending order of \(\abs{h_i - h_j}\)}
\KwIn{
  \begin{itemize}
    \item Eigenvalues \(\{h_0,\dots,h_{d-1}\}\),
    \item Matrices \(\mathbf{A}_{ij} \in \mathbb{C}^{|\mathcal{K}|\times (N_i + N_j)}\),
    \item An LP solver (equality constraints + nonnegative variables).
  \end{itemize}
}
\KwOut{
  \begin{itemize}
    \item A pair \((i^*, j^*)\) and $\vb*{p}_{0} \in \mathbb{R}^{N_{i}}_{\geq 0}$, $\vb*{p}_{1} \in \mathbb{R}^{N_{j}}_{\geq 0}$ with 
          \(\mathbf{A}_{i^* j^*}\,\vb*{p} = 0\), (where $\vb*{p}=(\vb*{p}_{0},\vb*{p}_{1})$),
    \item Or a report that no such solution exists.
  \end{itemize}
}

\BlankLine
\textbf{Step 1: Sort pairs \((i,j)\) in descending order of \(\lvert h_i - h_j\rvert\).}\;
\(\mathcal{P} \gets \{(i,j)\colon 0\le i,j \le d-1\}\);\;
\ForEach{$(i,j)\in\mathcal{P}$}{
  $D_{ij} \gets \lvert h_i - h_j\rvert$
}
$\mathcal{L} \gets \text{sort}\bigl(\mathcal{P}\bigr)\,\text{in descending order of}\,D_{ij}$\;

\BlankLine
\textbf{Step 2: For each \((i,j) \in \mathcal{L}\), solve for feasibility.}\;
\DontPrintSemicolon

\ForEach{$(i,j)\,\in\, \mathcal{L}$}{
  $\mathrm{Re}\!\bigl(\mathbf{A}_{ij}\bigr)\vb*{p}=0,\quad
   \mathrm{Im}\!\bigl(\mathbf{A}_{ij}\bigr)\vb*{p}=0,\quad
   \vb*{p}_{i}\ge 0,\quad
   \sum_{k=1}^{N_{i}}[\vb*{p}_{i}]_{k}=1,\quad \vb*{p}_{j}\ge 0 \quad \sum_{k=1}^{N_{j}}[\vb*{p}_{j}]_{k}=1.$ \tcp{ (to exclude $\vb*{p}=0$).}

  \If{LP is feasible}{
    \Return{$(i,j)$ and $\vb*{p}$}
  }
  \Else{
    \Continue
  }
}
\Return{\text{No feasible solution.}}

\end{algorithm}

\section{Sufficient conditions not satisfied}
\subsection{In the infinite $R$ limit}
In the main text, we discuss the potential challenges that arise when the sufficient condition in Theorem \ref{supple:theorem1} is not satisfied. In such cases, the signal Hamiltonian may induce logical errors that cannot be fully corrected by AutoQEC with finite $R$. In addition, in Sec. \ref{nsss}, we numerically study the cases when the sufficient condition is not satisfied.

In this section, we show that even if the sufficient condition does not hold, AutoQEC can restore the HS in the infinite $R$ limit ($R \to \infty$), provided that HNLS condition is satisfied. This result is formally stated in the following theorem:
\begin{theorem}\label{theorem22}
    In the infinite $R$ limit ($R \to \infty$), there exists an AutoQEC scheme that can restore the HS if and only if HNLS condition ($\hat{H} \notin \mathrm{span}\{\mathcal{K}^{[\sim 1]}\}$) is satisfied. In this case, the QFI of the AutoQEC state is given by
    \begin{align} 
        F[\hat{\rho}(t)]:=4t^{2}\left(\mathrm{Tr}[\hat{H}^{2}_{0}\hat{\rho}(0)]-\mathrm{Tr}[\hat{H}_{0}\hat{\rho}(0)]^{2}\right),
    \end{align}
    where $\hat{H}_{0}=\hat{\Pi}_{\mathcal{C}}\hat{H}\hat{\Pi}_{\mathcal{C}}$ and $\hat{\rho}(0) \in \mathcal{C}$ is the initial quantum probe.
\end{theorem}

\emph{Brief review of the ideal QEC.---} For the proof of Theorem \ref{theorem22}, we introduce the \textit{ideal quantum error correction} (QEC) scheme, originally proposed in Ref. \cite{qec-zhou2018achieving}. If the signal Hamiltonian satisfies HNLS condition, it can be decomposed as $\hat{H} = \hat{H}_{\perp} + \hat{H}_{\parallel}$, where $\hat{H}_{\perp}$ and $\hat{H}_{\parallel}$ are defined as follows. The component $\hat{H}_{\perp}$ satisfies the orthogonality condition $\mathrm{Tr}[\hat{H}_{\perp} \hat{K}] = 0$ for all $\hat{K} \in \mathrm{span} \{\mathcal{K}^{[\sim 1]}\}$, ensuring that it is orthogonal to the Lindblad span. Conversely, $\hat{H}_{\parallel}$ belongs to the Lindblad span, i.e., $\hat{H}_{\parallel} \in \mathrm{span} \{\mathcal{K}^{[\sim 1]}\}$.

Since the identity operator $\hat{I}$ is contained within the Lindblad span, i.e., $\hat{I} \in \mathrm{span}\{\mathcal{K}^{[\sim 1]}\}$, the component $\hat{H}_{\perp}$ must be traceless and Hermitian. Consequently, applying the spectral decomposition, $\hat{H}_{\perp}$ can be expressed as
\begin{align}
    \hat{H}_{\perp}=\frac{1}{2}\mathrm{Tr}[\vert{\hat{H}_{\perp}}\vert](\hat{\rho}_{0}-\hat{\rho}_{1})
\end{align}
where $\hat{\rho}_{0}$ and $\hat{\rho}_{1}$ are trace-one positive semidefinite matrices with mutually orthogonal support. Here, $|\hat{H}_{\perp}|$ is defined as the absolute value of $\hat{H}_{\perp}$, given by $|\hat{H}_{\perp}| = \sqrt{\hat{H}_{\perp}^{2}}$.

Next, we consider the QEC code $\mathcal{C}$ with noiseless ancilla that is chosen as a two-dimensional subspace of $\mathcal{H}_{p}\otimes \mathcal{H}_{\mathrm{A}}$ spanned by the code words $\{\ket{C_{0}},\ket{C_{1}}\}$, where the code words are normalized purification of $\hat{\rho}_{0}$ and $\hat{\rho}_{1}$ respectively, with orthogonal support in $\mathcal{H}_{\mathrm{A}}$. Here $\mathcal{H}_{p}$ and $\mathcal{H}_{\mathrm{A}}$ denote the Hilbert spaces of the physical qubit system and the noiseless ancilla system respectively. Since $\ket{C_{0}}$ and $\ket{C_{1}}$ have orthogonal support on $\mathcal{H}_{\mathrm{A}}$, for any $\hat{K}$ acting on $\mathcal{H}_{p}$ must satisfy the following orthogonality conditions:
\begin{align}
    \langle C_{0} \vert \hat{K}\otimes \hat{I}_{\mathrm{A}} \vert C_{1} \rangle=\langle C_{1} \vert \hat{K}\otimes \hat{I}_{\mathrm{A}} \vert C_{0} \rangle =0~,\forall~ \hat{K} \in \mathrm{span}\{\mathcal{K}^{[\sim 1]}\}. \label{KLcondition1}
\end{align}
Moreover, we note that $\mathrm{Tr}[\hat{H}_{\perp} \hat{K}]=0 ~,\forall~ \hat{K} \in \mathrm{span}\{\mathcal{K}^{[\sim 1]}\}$, which can be expressed as
\begin{align}
    \mathrm{Tr}[\hat{H}_{\perp} \hat{K}]=0= \frac{1}{2}\mathrm{Tr}\left[\abs{\hat{H}_{\perp}}\right]\mathrm{Tr}[(\hat{\rho}_{0}-\hat{\rho}_{1}) \hat{K}]= \frac{1}{2}\mathrm{Tr}\left[\abs{\hat{H}_{\perp}}\right] \left(\brao{C_{0}}\hat{K}\keto{C_{0}}-\brao{C_{1}}\hat{K}\keto{C_{1}}\right) ~,\forall~ \hat{K} \in \mathrm{span}\{\mathcal{K}^{[\sim 1]}\}.
\end{align}
It follows that
\begin{align}
    \langle C_{0} \vert \hat{K}\otimes \hat{I}_{\mathrm{A}} \vert C_{0} \rangle = \langle C_{1} \vert \hat{K}\otimes \hat{I}_{\mathrm{A}} \vert C_{1} \rangle ~,\forall~ \hat{K} \in \mathrm{span}\{\mathcal{K}^{[\sim 1]}\}.\label{KLcondition2}
\end{align}
Eqs. \eqref{KLcondition1} and \eqref{KLcondition2} ensure that the code $\mathcal{C}$ spanned with the codewords $\{\keto{C_{0}},\keto{C_{1}}\}$ satisfy the Knill-Laflamme condition for the error set $\mathcal{K}^{[\sim 1]}$.

Based on the chosen QEC code and the application of the ideal QEC, the error-corrected state at time $t$ is given by $\hat{\rho}(t)=e^{-iwt\hat{H}_{0}}\hat{\rho}(0)e^{iwt\hat{H}_{0}}$ where the effective signal Hamiltonian $\hat{H}_{0}$ is defined as 
\begin{align}
    \hat{H}_{0}:=\hat{\Pi}_{\mathcal{C}}\hat{H}\hat{\Pi}_{\mathcal{C}} = \langle C_{0} \vert \hat{H} \otimes \hat{I}_{\mathrm{A}} \vert C_{0} \rangle \dyado{C_{0}}+ \langle C_{1} \vert \hat{H} \otimes \hat{I}_{\mathrm{A}} \vert C_{1} \rangle \dyado{C_{1}}
\end{align}
satisfying $\langle C_{0} \vert \hat{H} \otimes \hat{I}_{\mathrm{A}} \vert C_{0} \rangle \neq  \langle C_{1} \vert \hat{H} \otimes \hat{I}_{\mathrm{A}} \vert C_{1} \rangle $. Here, $\hat{\Pi}_{\mathcal{C}}:=\dyado{C_{0}}+\dyado{C_{1}}$ is the projector onto the code $\mathcal{C}$.

\emph{AutoQEC with $R \to \infty$.---} Next, we consider AutoQEC in the infinite $R$ limit ($R\to\infty$). Let us prove the \textit{if} direction. The \textit{only if} direction can be readily established by applying the same analysis used in the ideal QEC setting of Ref.~\cite{qec-zhou2018achieving} of Methods section. Suppose that the signal Hamiltonian $\hat{H}$ satisfies HNLS condition. For the AutoQEC, we employ the same code that is spanned by the code words $\{\keto{C_{0}},\keto{C_{1}}\}$. Furthermore, we incorporate the engineered dissipation defined in Eq. \eqref{supple:engineereddissi} to implement AutoQEC. For simplicity, in the remainder of the paper, we omit $\hat{I}_{\mathrm{A}}$ in expressions of the form $\hat{\Omega} \otimes \hat{I}_{\mathrm{A}}$ where $\hat{I}_{\mathrm{A}}$ denotes the identity operator acting on the noiseless ancilla mode.

The Lindblad equation governing AutoQEC dynamics is given by
\begin{align}
    \frac{d\hat{\rho}(t)}{dt}= -iw[\hat{H},\hat{\rho}(t)]+\kappa\tilde{\mathcal{L}}_{\mathrm{n}}[\hat{\rho}(t)]+R\kappa\tilde{\mathcal{L}}_{\mathrm{E}}[\hat{\rho}(t)]. \label{Aqeclind}
\end{align}
Equivalently, it can be expressed as
\begin{align}
    \hat{\rho}(t+\delta t)=\hat{\rho}(t)+ \delta t\left(-iw [\hat{H},\hat{\rho}(t)]+\kappa\tilde{\mathcal{L}}_{\mathrm{n}}[\hat{\rho}(t)]+R\kappa\tilde{\mathcal{L}}_{\mathrm{E}}[\hat{\rho}(t)]\right) + O(\delta t ^{2}), \label{lindvariy}
\end{align}
for infinitesimal $\delta t$ limit ($\delta t \to 0$).
Since we are considering an infinite $R$ limit, let us consider the following ansatz for the density matrix expansion of the Lindblad equation:
\begin{align}
    \hat{\rho}(t)=\hat{\rho}^{(0)}(t)+\frac{1}{R}\hat{\rho}^{(1)}(t)+\frac{1}{R^{2}}\hat{\rho}^{(2)}(t) + \cdots. \label{ansatz}
\end{align}
Here, we further assume that the initial state is in the code space and does not depends on $R$, i.e., $\hat{\rho}(0)=\hat{\rho}^{(0)}(0) \in \mathcal{C}$. In the infinite $R$ limit, $\hat{\rho}(t)=\hat{\rho}^{(0)}(t)$ for all $t\geq 0$.
Notably, due to the completely positive trace-preserving (CPTP) condition, $\hat{\rho}(t)$ cannot contain terms that scale polynomially with $R$. Next, let us put the ansatz (in Eq. \eqref{ansatz}) into the Lindblad equation (in Eq. \eqref{Aqeclind}) and consider the term related to the first order of $R$ that is
\begin{align}
    \text{$1$st order}: \quad R\kappa\tilde{\mathcal{L}}_{\mathrm{E}}[\hat{\rho}^{(0)}(t)]=0. \label{1storder}
\end{align}
Eq. \eqref{1storder} implies
\begin{align}
    \tilde{\mathcal{L}}_{\mathrm{E}}[\hat{\rho}^{(0)}(t)]=\tilde{\mathcal{L}}_{\mathrm{E}}[\hat{\rho}(t)]=0, \quad \forall~ t \geq 0, \label{darkspace}
\end{align}
in the infinite $R$ limit. Here, we note that, from the definition of the engineered dissipation defined in Eq. \eqref{supple:engineereddissi}, one can easily prove that Eq. \eqref{darkspace} is satisfied if and only if $\hat{\rho}(t) \in \mathcal{C}$.

Since $\hat{\rho}(t)$ satisfies Eq. \eqref{darkspace}, it must also satisfy the following condition:
\begin{align}
    \tilde{\mathcal{P}}_{\mathrm{E}}\left[\hat{\rho}(t)\right] = \hat{\rho}(t)~\forall~ t \geq 0,
\end{align}
where $\tilde{\mathcal{P}}_{\mathrm{E}}:= \lim_{u \to \infty}e^{\tilde{\mathcal{L}}_{\mathrm{E}}u}$ is CPTP projector.
As a consequence, applying $\tilde{\mathcal{P}}$ to both sides of Eq. \eqref{lindvariy} yields:
\begin{align}
    \tilde{\mathcal{P}}_{\mathrm{E}}\left[\hat{\rho}(t+\delta t)\right]=\tilde{\mathcal{P}}_{\mathrm{E}}\left[\hat{\rho}(t)\right]+\delta t\tilde{\mathcal{P}}_{\mathrm{E}}\left[ -iw [\hat{H},\hat{\rho}(t)]+\kappa\tilde{\mathcal{L}}_{\mathrm{n}}[\hat{\rho}(t)]\right] + O(\delta t ^{2}).  \label{peapplied}
\end{align}
Here, we note that the CPTP projector $\tilde{\mathcal{P}}_{\mathrm{E}}$ satisfy the following two properties \cite{aqec-lebreuilly2021autonomousquantumerrorcorrection}:
(1) $\tilde{\mathcal{P}}_{\mathrm{E}}\tilde{\mathcal{L}}_{\mathrm{n}}[\hat{\rho}(t)]=0$ for $\hat{\rho}(t) \in \mathcal{C}$. (2) Analytical expression of $\tilde{\mathcal{P}}_{\mathrm{E}}$ is given by \cite{aqec-lebreuilly2021autonomousquantumerrorcorrection} 
\begin{align}
    \tilde{\mathcal{P}}_{\mathrm{E}}[\hat{\rho}]= \hat{\Pi}_{\mathcal{C}}\hat{\rho}\hat{\Pi}_{\mathcal{C}}+ \sum_{n=1}^{c}\sum_{i_{n}=1}^{p_{n}}\hat{L}^{[n]}_{\mathrm{E},i_{n}}\hat{\rho}\hat{L}^{[n]\dagger}_{\mathrm{E},i_{n}} + \sum_{q=1}^{q_{\mathrm{max}}} \hat{L}^{[\mathrm{res}]}_{\mathrm{E},q}\hat{\rho}\hat{L}^{[\mathrm{res}]\dagger}_{\mathrm{E},q}.
\end{align} 
Furthermore, exploiting the fact that $\tilde{\mathcal{P}}_{\mathrm{E}}\tilde{\mathcal{L}}_{\mathrm{n}}\hat{\Pi}_{\mathcal{C}}=0$ \cite{aqec-lebreuilly2021autonomousquantumerrorcorrection}, Eq. \eqref{peapplied} can be reformulated as
\begin{align}
    \hat{\rho}(t+\delta t)=\hat{\rho}(t)+\delta t\left( -iw [\hat{H}_{0},\hat{\rho}(t)]\right)+ O(\delta t ^{2}).  
\end{align}
where $\hat{H}_{0}:=\hat{\Pi}_{\mathcal{C}} \hat{H} \hat{\Pi}_{\mathcal{C}}$. Therefore, the AutoQEC state with infinite $R$ is $\hat{\rho}(t)=e^{-i\hat{H}_{0}t}\hat{\rho}(0)e^{i\hat{H}_{0}t}$, whose QFI is $F[\hat{\rho}(t)]:=4t^{2}\left(\mathrm{Tr}[\hat{H}^{2}_{0}\hat{\rho}(0)]-\mathrm{Tr}[\hat{H}_{0}\hat{\rho}(0)]^{2}\right)$, which shows the HS.

We conclude by numerically examining two representative scenarios when $R$ is sufficiently large. In both cases, the Knill-Laflamme condition is satisfied, ensuring correctability of the noise. The key distinction lies in the HNLS condition: it is satisfied in the first scenario and violated in the second. In both cases, we consider the signal Hamiltonian as $\hat{H}=\hat{Z}_{1}$, and we assume access to a noiseless ancilla qubit.
In the first scenario, the HNLS condition is satisfied. Specifically, We analyze a local bit flip noise model with a single Lindblad operator $\hat{L}_{1}=\hat{X}_{1}$, where $\hat{X}_{1}$ denotes the Pauli-X operator acting on the first qubit. As a candidate code, we employ a two-qubit code with codewords $\keto{\mu_{0}}=\ket{0}\ket{0}_{\mathrm{A}}$ and $ \keto{\mu_{1}}=\ket{1}\ket{1}_{\mathrm{A}}$ where $\mathrm{A}$ denotes the noiseless ancilla qubit. It is straightforward to verify that both the Knill-Laflamme and HNLS conditions are satisfied in this setting.
In the second scenario, the HNLS condition is violated. We consider a local dephasing noise model with a single Lindblad operator $\hat{L}_{1}=\hat{Z}_{1}$, where $\hat{Z}_{1}$ denotes the Pauli-Z operator acting on the first qubit. As a candidate code, we use a two-qubit code with codewords $\keto{\mu_{0}}=\ket{+}\ket{+}_{\mathrm{A}}$ and $ \keto{\mu_{1}}=\ket{-}\ket{-}_{\mathrm{A}}$. Although the Knill-Laflamme condition is satisfied in this case, the HNLS condition is clearly violated, as the Hamiltonian is entirely contained within the Lindblad span.

\begin{figure}[t]
    \centering
    \includegraphics[width=0.65\linewidth]{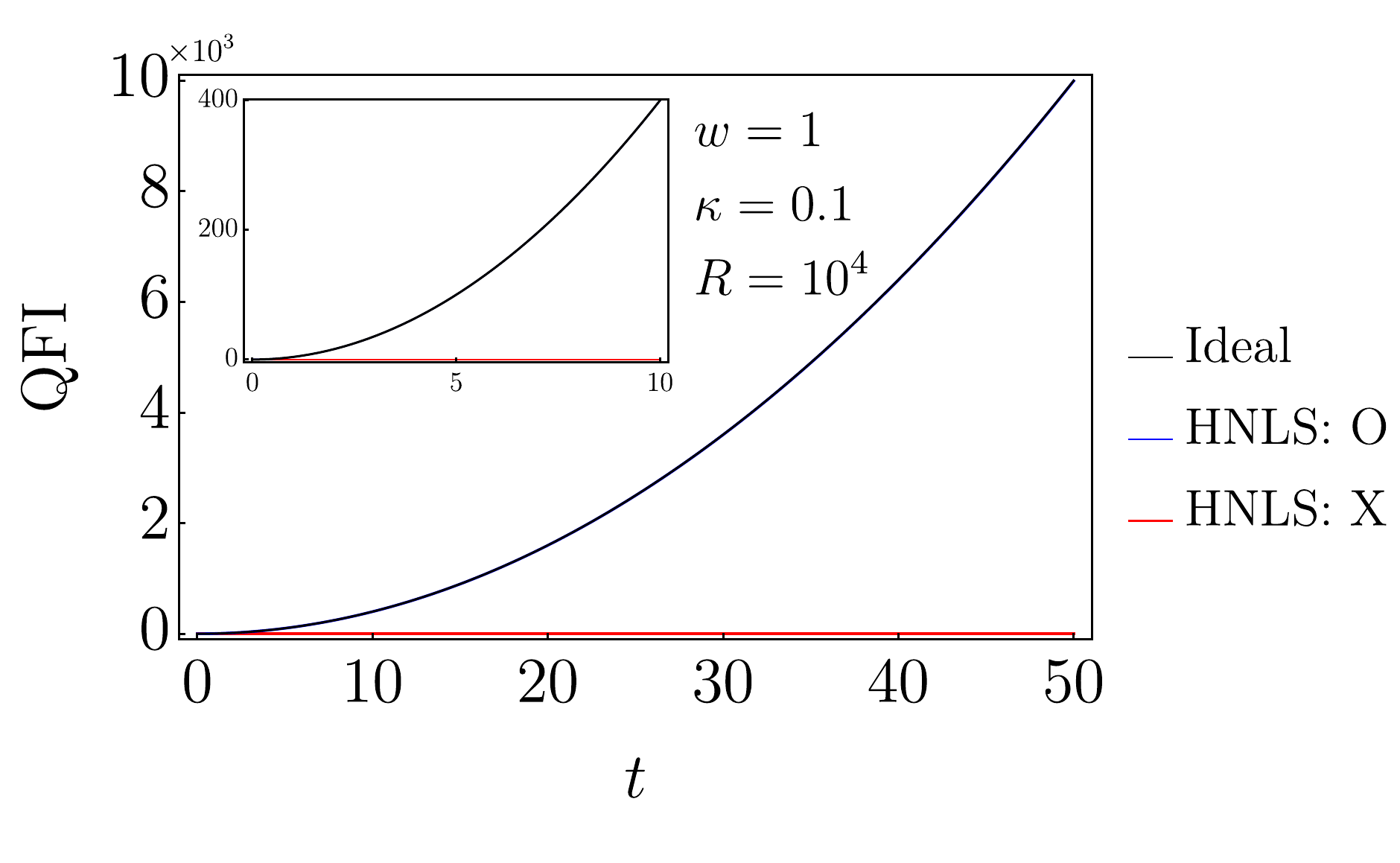}
    \caption{
    QFI as a function of sensing time $t$, with $R=10^{4}$, $w=1$, $\kappa=0.1$, and AutoQEC order $c=1$. The black line (labeled ``Ideal") corresponds to the noiseless case. The blue line (labeled ``HNLS: O") represent the QFI when both the Knill-Laflamme condition and HNLS condition are satisfied. We note that the blue line is nearly equal to the black line. The red line (labeled ``HNLS: X") shows the QFI when only the Knill-Laflamme condition is satisfied, while HNLS condition is violated. The inset shows a zoomed-in view of the QFI for a smaller range of $t$.
    }
    \label{fig:HNLSNotSatisfied}
\end{figure}

As shown in Fig.~\ref{fig:HNLSNotSatisfied}, although the Knill-Laflamme condition is satisfied in both scenarios, it is the HNLS condition that determines whether HS can be achieved. When the HNLS condition is satisfied, the QFI under AutoQEC—depicted by the blue line—closely matches that of the noiseless evolution, indicating near-optimal metrological performance. In contrast, when the HNLS condition is violated, the QFI—shown by the red line—remains close to zero for all evolution times $t$, despite the error correction being in place. This behavior illustrates that while the engineered dissipation effectively corrects the noise, it also suppresses the signal component encoded by the Hamiltonian if it lies within the Lindblad span. As a result, the metrological advantage is entirely lost.

\subsection{Numerical simulations: sufficient condition is not satisfied with finite $R$}\label{nsss}
In this section, we numerically study the QFI of the AutoQEC (with finite $R$) state under three different scenarios: (1) Both (P1) and (P2) are satisfied. (2) (P1) is violated while (P2) remains satisfied. (3) (P2) is violated while (P1) remains satisfied. For all the different scenarios, we consider $3$-qubit system with $w=1$, $\kappa=0.1$, and $c=1$.

\emph{Both (P1) and (P2) are satisfied.---} For scenario (1), we consider the same task as in the second numerical analysis presented in the main text. Specifically, we analyze the case where the signal Hamiltonian is given by $\hat{H}=\prod_{i=1}^{N}\hat{Z}_{i}$ and assume the presence of local dephasing noise. In addition, we consider the following code words
\begin{align}
    &\keto{\mu_{0}}=(\keto{+}^{\otimes 3}+\keto{-}^{\otimes 3})/\sqrt{2},\\
    &\keto{\mu_{1}}=(\keto{+}^{\otimes 3}-\keto{-}^{\otimes 3})/\sqrt{2}.
\end{align}
In this case, the sufficient condition in Theorem \ref{supple:theorem1} is satisfied up to AutoQEC order $c=1$, which implies that both (P1) and (P2) are satisfied.

\emph{(P1) is violated while (P2) remains satisfied.---} For scenario (2), we examine the case where the signal Hamiltonian is given by $\hat{H}=\hat{Z}_{1}\hat{Z}_{2}\hat{Z}_{3}$ and assume the presence of local Pauli-$X$ noise, where the corresponding sensing dynamics (in the absence of AutoQEC) are governed by the following Lindblad master equation
\begin{align}
    \frac{d\hat{\rho}}{dt}= -iw[\hat{H},\hat{\rho}(t)]+\kappa \sum_{i=1}^{3}\hat{X}_{i}\hat{\rho}\hat{X}_{i}. 
\end{align}
Notably, the error operators $\hat{X}_{i}$ do not commute with $\hat{H}$ for any $i$, indicating that (P1) is not satisfied.
Next, we consider the following code words
\begin{align}
    \keto{\mu_{0}}= \keto{0}^{\otimes 3}, \quad \keto{\mu_{1}}= \keto{1}^{\otimes 3}.
\end{align}
In this case, the code satisfies the Knill-Laflamme condition for the error set $\mathcal{E}^{[\sim 1]}=\{\hat{I},\hat{X}_{1},\hat{X}_{2},\hat{X}_{3}\}$.
Based on the code words, the signal Hamiltonian can be expressed as
\begin{align}
    \hat{H}=\sum_{j=0}^{1}\sum_{n=0}^{1}\sum_{i_{n}=1}^{3} (-1)^{j+n} \keto{ \mu^{[n]}_{j,i_{n}} } \brao{  \mu^{[n]}_{j,i_{n}} }, 
\end{align}
where $\keto{\mu^{[1]}_{j,i_{n}}} := \hat{X}_{i_{n}}\keto{\mu_{j}}$. It follows that the signal Hamiltonian satisfies (P2).

\emph{(P2) is violated while (P1) remains satisfied.---} For scenario (3), we analyze the case where the signal Hamiltonian is given by $\hat{H}=\frac{1}{2}\left(\hat{Z}_{1}\hat{Z}_{2}\hat{Z}_{3}+\sum_{k=1}^{3}\hat{Z}_{i}\right)$. Furthermore, we assume that local dephasing noise is present in the sensing dynamics. We note that $\hat{H}$ commutes with $\hat{Z}_{i}$ for all $i$, indicating that (P1) is satisfied. We then consider the following code words
\begin{align}
    \keto{\mu_{0}}=(\keto{+}^{\otimes N}+\keto{-}^{\otimes N})/\sqrt{2},\quad \keto{\mu_{1}}=(\keto{+}^{\otimes N}-\keto{-}^{\otimes N})/\sqrt{2}.
\end{align}
In this case, the code satisfies the Knill-Laflamme condition for the error set $\mathcal{E}^{[\sim 1]}=\{\hat{I},\hat{Z}_{1},\hat{Z}_{2},\hat{Z}_{3}\}$. Based on the code words, the signal Hamiltonian can be expressed as
\begin{align}
    \hat{H}=\frac{1}{2}\sum_{j=0}^{1}\sum_{n=0}^{1}\sum_{i_{n}=1}^{3} (-1)^{j}\dyado{\mu^{[n]}_{j,i_{n}}} + \frac{1}{2}\sum_{j=0}^{1}\sum_{n=0}^{1}\sum_{i_{n}=1}^{3} \left(\keto{\mu^{[n]}_{j,i_{n}}}\brao{\mu_{j}}+\keto{\mu_{j}}\brao{\mu^{[n]}_{j,i_{n}}}\right)
\end{align}
where $\keto{\mu^{[1]}_{j,i_{n}}} := \hat{Z}_{i_{n}}\keto{\mu_{j}}$. Here, one can easily find that the signal Hamiltonian does not satisfy (P2).

\begin{figure}[t]
    \centering
    \includegraphics[width=0.95\linewidth]{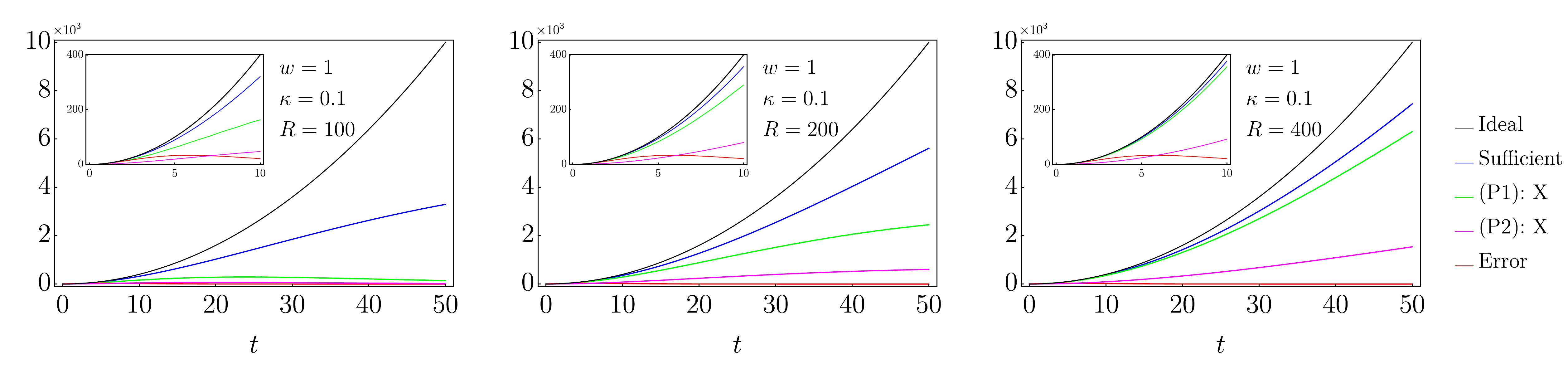}
    \caption{QFI as a function of sensing time $t$, different values of $R$, with $w=1$, $\kappa=0.1$, and AutoQEC order $c=1$. The black line (labeled ``Ideal") and the red line (labeled ``Error") represent the QFI of the noiseless case and without AutoQEC case respectively. The green line (labeled ``(P1): X") and the pink line (labeled ``(P2): X") represent the QFI when (P1) and (P2) are not satisfied respectively. The inset shows a zoomed-in view of the QFI for a smaller range of $t$.
    }
    \label{fig:SuffiX}
\end{figure}

For all scenarios, we exploit $\keto{\mu_{0}}+\keto{\mu_{1}}$ as a initial quantum probe where the ideal QFI becomes $F[\hat{\rho}_{\mathrm{id}}(t)]=4t^{2}$. We then numerically simulate the three different scenarios (1)-(3) with $R=100$, $R=200$, and $R=400$. See Fig. \ref{fig:SuffiX}. In the numerical simulations, we find that, as discussed in the main text, if either (P1) or (P2) is violated, the performance of AutoQEC is degraded compared to the case where both (P1) and (P2) are satisfied---equivalently, the case where the sufficient condition in Theorem \ref{supple:theorem1} is satisfied. In particular, when (P2) is violated, the signal Hamiltonian accumulates errors faster than the natural dissipation, and AutoQEC exhibits the worst performance among the three scenarios (1)–(3).

\section{Phase estimations satisfying the sufficient condition }
\subsection{$\hat{H}=\sum_{i=1}^{N}\hat{Z}_{i}$ under correlated dephasing noise}
We consider $N$-qubit canonical phase estimation where the signal Hamiltonian is given by $\hat{H}=\sum_{i=1}^{N}\hat{Z}_{i}$, in the presence of the correlated dephasing noise. We note that (T1) is always satisfied in this scenario. The sensing dynamics (in the absence of AutoQEC) are governed by the following Lindblad master equation:
\begin{align}
    \frac{d\hat{\rho}}{dt}= -iw[\hat{H},\hat{\rho}(t)]+\kappa \sum_{i,j=1}^{N}[\mathbf{C}]_{ij}\left(\hat{Z}_{i}\hat{\rho}\hat{Z}_{j}-\frac{1}{2}\{\hat{Z}_{i}\hat{Z}_{j},\hat{\rho}\}\right), \label{lindnume}
\end{align}
where $\hat{Z}_{i}$ represents the Pauli-$Z$ operator acting on $i$th qubit. $\mathbf{C}$ is the correlation matrix that characterizes the spatial structure of the noise. Specifically, the off-diagonal elements $[\mathbf{C}]_{ij}$ quantify the noise correlation between $i$th qubit and $j$th qubit. 
Equivalently, the correlated dephasing noise can be expressed as
\begin{align}
    \frac{d\hat{\rho}}{dt}= -iw[\hat{H},\hat{\rho}(t)]+\kappa \sum_{i=1}^{N}\left(\hat{L}_{i}\hat{\rho}\hat{L}_{i}-\frac{1}{2}\{\hat{L}_{i}\hat{L}_{i},\hat{\rho}\}\right), \label{usingD}
\end{align}
where the Lindblad operators are defined as $\hat{L}_{i}=\sum_{j=1}^{N}[\mathbf{D}]_{ij}\hat{Z}_{j}$ for all $1\leq i\leq N$. Here, the $\mathbf{D}$ is $N \times N$ real matrix that satisfies $\mathbf{D}^{\mathrm{T}}\mathbf{D}=\mathbf{C}$.

Next, we classify the correlated dephasing noise into three distinct cases based on the structure of $\mathbf{D}$. The first case is when the row vectors of $\mathbf{D}$, denoted as $\{[\mathbf{D}]_{i}:=\{[\mathbf{D}]_{i1},[\mathbf{D}]_{i2},\cdots,[\mathbf{D}]_{iN}\}\}_{i=1}^{N}$, form a basis set for $N$-dimensional real space. In other words, $\mathbf{D}$ is full-rank matrix, equivalently, $\mathbf{C}$ is full-rank matrix. In this case, there always exists a set of constants $\{c_{i} \in \mathbb{R}\}_{i=1}^{N}$ such that $\sum_{i=1}^{N} c_{i}[\mathbf{D}]_{ij}\hat{Z}_{j}=\sum_{i=1}^{N}\hat{Z}_{i}$, which implies the violation of HNLS condition. Consequently, AutoQEC cannot restore the HS. The second case is when $\sum_{j=1}^{N}[\mathbf{D}]_{ij}=0$ for all $i$. In this scenario, the HS can be achieved without applying an error correction. This can be demonstrated by employing the quantum probe $\keto{\psi_{0}}=\keto{0}^{\otimes N}+\keto{1}^{\otimes N}$. Under this choice of the probe, the signal state $\keto{\psi(t)}:=e^{-i\hat{H}wt}\keto{\psi_{0}}$ remains unaffected by noise, specifically, the signal state satisfies $\tilde{\mathcal{L}}_{\mathrm{n}}[\dyado{\psi(t)}]=0$, indicating that the noise does not degrade the estimation precision. As a result, the QFI is given by $F[\dyado{\psi(t)}]=4N^{2}t^{2}$, demonstrating the achievement of HS. The last case is when $[\mathbf{D}]_{ij}$ does not satisfy the conditions of the previous two cases. In our numerical simulation, we mainly focus on this case. Notably, in this case, the ideal QEC has been shown to restore HS \cite{qec-zhou2018achieving, qec-PhysRevLett.122.040502}. However, for AutoQEC, the validity of the sufficient condition in Theorem \ref{supple:theorem1} is not guaranteed and depends on the structure of the correlation matrix $\mathbf{C}$. For example, in the case of $3$-qubit system with the correlation matrix
\begin{align}
    \mathbf{C}=\begin{pmatrix}
       8 & 6 & 4\\
        6 & 6 & 6 \\
        4 & 6 & 8 \\
    \end{pmatrix},~~\text{or equivalently,~~}\mathbf{D}=\begin{pmatrix}
       2 & 1 & 0\\
        0 & 1 & 2 \\
        2 & 2 & 2 \\
    \end{pmatrix},
\end{align}
there exist no $h_{i}$, $h_{j \neq i}$, $\vb*{p}_{i}$, and $\vb*{p}_{j \neq i}$ satisfying (T2).

For the first numerical simulation in the main text, we consider $3$-qubit system with $w=1$, $\kappa=0.1$, and the correlation matrix
\begin{align}
    \mathbf{C}=\frac{1}{10}\begin{pmatrix}
       16 & -4 & -4\\
        -4 & 7 & -5 \\
        -4 & -5 & 7 \\
    \end{pmatrix},~~\text{or equivalently,~~}\mathbf{D}=\begin{pmatrix}
       \frac{2}{\sqrt{5}} & -\frac{1}{\sqrt{5}} & 0\\
        0 & \frac{1}{\sqrt{2}} & -\frac{1}{\sqrt{2}} \\
        \frac{2}{\sqrt{5}} & 0 & -\frac{1}{\sqrt{5}} \\
    \end{pmatrix}.
\end{align}
For $3$-qubit case, the signal Hamiltonian can be expressed as
\begin{align}
    \hat{H}=3\dyado{000}-3\dyado{111}+h_{0}\sum_{k=1}^{3}\dyado{h^{(k)}_{0}}+h_{1}\sum_{k=1}^{3}\dyado{h^{(1)}_{0}}
\end{align}
where $h_{0}=1$, $h_{1}=-1$, $\{\keto{h^{(1)}_{0}},\keto{h^{(2)}_{0}},\keto{h^{(3)}_{0}}\}=\{\keto{100},\keto{010},\keto{001}\}$, and  $\{\keto{h^{(1)}_{1}},\keto{h^{(2)}_{1}},\keto{h^{(3)}_{1}}\}=\{\keto{011},\keto{101},\keto{110}\}$. In this scenario, the sufficient condition in Theorem \ref{supple:theorem1} is satisfied up to AutoQEC order $c=1$. More specifically, $h_{0}$, $h_{1}$, $\vb*{p}_{0}=\frac{1}{10}\{4,3,3\}$, and $\vb*{p}_{1}=\frac{1}{10}\{4,3,3\}$ satisfying (T2). Therefore, based on Eq. \eqref{codewords}, we consider the following code words:
\begin{align}
    \keto{\mu_{0}}=\sqrt{\frac{4}{10}}\keto{100}+\sqrt{\frac{3}{10}}\keto{010}+\sqrt{\frac{3}{10}}\keto{001},~~\keto{\mu_{1}}=\sqrt{\frac{4}{10}}\keto{011}+\sqrt{\frac{3}{10}}\keto{101}+\sqrt{\frac{3}{10}}\keto{110}.
\end{align}
Consequently, applying the Gram-Schmidt orthogonalization, we obtain the basis vectors of the first order correctable error space:
\begin{align}
    &\keto{\mu^{[1]}_{0,1}}=-\sqrt{\frac{12}{22}}\keto{100}+\sqrt{\frac{9}{22}}\keto{010}+\sqrt{\frac{1}{22}}\keto{001},~~\keto{\mu^{[1]}_{0,2}}=-\sqrt{\frac{3}{55}}\keto{100}-\sqrt{\frac{16}{55}}\keto{010}+\sqrt{\frac{36}{55}}\keto{001},\\
    &\keto{\mu^{[1]}_{1,1}}=\sqrt{\frac{12}{22}}\keto{011}-\sqrt{\frac{9}{22}}\keto{101}-\sqrt{\frac{1}{22}}\keto{110},~~\keto{\mu^{[1]}_{1,2}}=\sqrt{\frac{3}{55}}\keto{011}+\sqrt{\frac{16}{55}}\keto{101}-\sqrt{\frac{36}{55}}\keto{110}.
\end{align}
Furthermore, the residual space is spanned by $\{\keto{\phi_{1}},\keto{\phi_{2}}\}:= \{\keto{000},\keto{111}\}$. We note that the choice of $\keto{\Phi_{q}}$ does not affect the performance of AutoQEC, since the AutoQEC state never transitions into the residual space throughout its evolution.

\subsection{$\hat{H}=\prod_{i=1}^{N}\hat{Z}_{i}$ under local dephasing noise}\label{supplesec:localdeph}
We analyze $N$-qubit phase estimation where the signal Hamiltonian is given by $\hat{H}=\prod_{i=1}^{N}\hat{Z}_{i}$, in the presence of the correlated dephasing noise. In this setting, the conditions (C1) and (C3) in Corollary \ref{corollary2} are always satisfied in this case. The sensing dynamics are governed by the Lindblad master equation in Eq. \eqref{lindnume}. The error set $\mathcal{E}^{[\sim \lfloor{\frac{N-1}{2}}\rfloor]}$ consists of at most $\lfloor{\frac{N-1}{2}}\rfloor$ numbers of the product of the Pauli-$Z$ operators which implies that $\mathcal{K}^{[\sim \lfloor{\frac{N-1}{2}}\rfloor]}$ consists of at most $2\lfloor{\frac{N-1}{2}}\rfloor < N$ numbers of the product of the Pauli-$Z$ operators. As a result, all the elements of $\mathcal{K}^{[\sim \lfloor{\frac{N-1}{2}}\rfloor]}$ are orthogonal to the signal Hamiltonian, which ensures the satisfaction of (C2). Therefore, according to Corollary \ref{corollary2}, AutoQEC can be performed up to order $c$ for all $1 \leq c \leq \lfloor{\frac{N-1}{2}}\rfloor$ using the code spanned by the code words
\begin{align}
    \keto{\mu_{0}}=\frac{1}{\sqrt{2}}(\keto{+}^{\otimes N}+\keto{-}^{\otimes N}),~~\keto{\mu_{1}}=\frac{1}{\sqrt{2}}(\keto{+}^{\otimes N}-\keto{-}^{\otimes N}).
\end{align}
For the numerical simulation, we consider $5$-qubit system with $w=1$, $\kappa=0.1$, and $R=100$ in the presence of the local dephasing noise where the correlation matrix is the identity matrix. As we inspected above, AutoQEC can be performed up to order $c=1,2$. The basis vectors of the first-order correctable error spaces are given by
\begin{align}
    \{\keto{\mu^{[1]}_{\alpha, i_{n}}}\}_{i_{n}=1}^{5}=\{\hat{Z}_{i_{n}} \keto{\mu_{\alpha}} \}_{i_{n}=1}^{5},~~\alpha \in \{0,1\},
\end{align}
and the basis vectors of the second-order correctable error spaces are given by
\begin{align}
    \{\keto{\mu^{[2]}_{\alpha, i_{n}}}\}_{i_{n}=1}^{10}=\{\hat{Z}_{a}\hat{Z}_{b} \keto{\mu_{\alpha}} \vert a>b,~1 \leq b \leq N-1,~ 2 \leq a \leq N \},~~\alpha \in \{0,1\}.
\end{align}

\emph{$c=2$ case.---} We first consider the AutoQEC order $c=2$. According to Corollary \ref{corollary2}, AutoQEC up to order $c=2$ can be performed by applying the engineered dissipation defined in Eq. \eqref{supple:engineereddissi}. We note that in this case, there is no residual space since the union of the code space, the first-order correctable error space, and the second-order correctable error space is the total Hilbert space of the $5$-qubit system. Therefore, the second term in Eq. \eqref{supple:engineereddissi} $\sum_{q=1}^{q_{\mathrm{max}}}D[\hat{L}^{[\mathrm{res}]}_{\mathrm{E},q}]$ is $0$.

\emph{$c=1$ case.---} Next, we consider AutoQEC order $c=1$. In this case, AutoQEC can also be performed by applying the engineered dissipation in Eq. \eqref{supple:engineereddissi}. For $c=1$, there is a residual space, and the basis of the residual space is given by
\begin{align}
    \{\keto{\phi_{q}}\}_{q=1}^{20}=  \{\keto{\mu^{[2]}_{0, i_{n}}}\}_{i_{n}=1}^{10} \cup \{\keto{\mu^{[2]}_{1, i_{n}}}\}_{i_{n}=1}^{10}.
\end{align}
For the numerical simulation, we set $\keto{\Phi_{q}}$ in $\hat{L}^{[\mathrm{res}]}_{\mathrm{E},q}=\keto{\Phi_{q}}\brao{\phi_{q}}$ as $\keto{\Phi_{q}}=\frac{1}{\sqrt{2}}(\keto{\mu_{0}}+\keto{\mu_{1}})$ for all $1 \leq q \leq 20$.

\section{Relevant complexity analysis on higher-order AutoQEC}

\subsection{Scaling of the recovery-jump count and the role of higher-order AutoQEC}
In the canonical construction of Lemma~1, the engineered dissipation comprises recovery jump operator $\hat{L}^{[n]}_{\mathrm{E},i_{n}}=\sum_{j=0}^{d_{\mathcal{C}}}\keto{\mu_{j}}\brao{\mu^{[n]}_{j,i_{n}}}$ for each orthonormal basis vector of the $n$-th correctable error space. Accordingly, correcting $n$-th order errors requires a set of recovery operators $\left\{\hat{L}^{[n]}_{\mathrm{E},i_{n}}\right\}_{i=1}^{p_{n}}$ to correct $n$-th order errors. Extending this construction to achieve AutoQEC up to order $c$, the total number of engineered recovery jump operators is given by
\begin{align}
N_{\mathrm{rec}}^{\mathrm{can}}
:=
\sum_{n=1}^{c} p_n,
\end{align}
where $p_n$ denotes the dimension of the $n$-th correctable error space per logical basis state.

Next, we now derive an upper bound on $N_{\mathrm{rec}}^{\mathrm{can}}$. To understand how $N_{\mathrm{rec}}^{\mathrm{can}}$ scales with the AutoQEC order $c$, it is useful to examine the recursive construction of the higher-order error sets. Let
\begin{align}
e_n := |\mathcal{E}^{[n]}|
\end{align}
denote the number of $n$-th order error operators. By construction, each $(n-1)$-th order error operator can generate up to $N_{\mathrm{n}}$ new error operators of the form $L_{\mathrm{n},a}E_{\ell}^{[n-1]}$, where $N_\mathrm{n}$ is the number of natural Lindblad operators, while each $(n-2)$-th order error operator generates one additional operator of the form $B E_{\ell}^{[n-2]}$ where $B=\sum_{a=1}^{{N}_{\mathrm{n}}}\hat{L}^{\dagger}_{\mathrm{n},a}\hat{L}_{\mathrm{n},a}$. Here we note that $\hat{E}^{[n]}\in \mathcal{E}^{[n]}$ by the definition in Eq. (5) in the main text. Therefore, in the worst case one has the recursive bound
\begin{align}
e_n \le N_\mathrm{n} e_{n-1} + e_{n-2}.\label{supple:recursion}
\end{align}
Moreover, since the $n$-th correctable error-space basis is obtained by Gram--Schmidt orthogonalization of the family generated by the code word $|\mu_i\rangle$ and the error states $\hat E_a^{[n]}|\mu_i\rangle$, one has
\begin{align}
p_n \le  e_n.
\end{align}
Hence, the growth of the number of recovery jump operators is controlled by the growth of the higher-order error sets.

In particular, the characteristic equation of the recursion in Eq. \eqref{supple:recursion} is
\begin{align}
r^2 - N_\mathrm{n} r - 1 = 0,
\end{align}
whose larger root is
\begin{align}
\lambda_+ = \frac{N_\mathrm{n} + \sqrt{N_\mathrm{n}^2+4}}{2}.
\end{align}
Thus, in the worst case, the number of higher-order error operators can scale as
\begin{align}
e_n = O(\lambda_+^n),
\end{align}
and therefore the number of required recovery jump operators can also grow exponentially with the AutoQEC order $c$ as
\begin{align}
    N_{\mathrm{rec}}^{\mathrm{can}} \le \sum_{n=1}^{c}e_{n}=O(\lambda_{+}^{c}).
\end{align}
We emphasize, however, that this bound represents a worst-case, model-independent estimate. Furthermore, the AutoQEC order $c$ need not be large in practice. Indeed, our numerical results indicate that even a modest choice, $c=2$, already yields substantial performance improvement. Finally, we examine the maximum achievable order $c$. The total dimension of the code space together with all correctable error spaces must fit within the finite-dimensional Hilbert space, which imposes the constraint that $c$ must satisfy
\begin{align}
d_{\mathcal{C}}\sum_{i=0}^{c}p_{n} \le d_{\mathcal{H}},
\end{align}
where $d_{\mathcal{H}}$ denotes the total dimension of the Hilbert space. As an illustrative example, consider the 5-qubit correlated-dephasing model under the repetition code discussed in Sec.~\ref{supplesec:localdeph}. In this case, one has $p_1=5$ and $p_2=10$. Therefore, the canonical construction uses
\begin{align}
N_{\mathrm{rec}}^{\mathrm{can}} = 5
\end{align}
for $c=1$, whereas for $c=2$ it uses
\begin{align}
N_{\mathrm{rec}}^{\mathrm{can}} = 5+10 = 15.
\end{align}
Moreover, for an $N$-qubit repetition code, AutoQEC can be implemented at most $1 \le c \le \lfloor (N-1)/2 \rfloor$. 

Importantly, the possible increase in the number of recovery jump operators should be understood together with the main advantage of higher-order AutoQEC. According to Theorem~1, the logical error scales as
\begin{align}
\epsilon = O(\kappa T/R^c),
\end{align}
so a larger AutoQEC order $c$ allows the same target accuracy to be achieved with a smaller engineered-dissipation strength $R$, or equivalently yields better protection for the same $R$. In this sense, higher-order AutoQEC provides a useful tradeoff: it can substantially relax the required dissipation strength, even though it may require more recovery jump operators. This tradeoff is especially relevant in metrological settings, where finite $R$ is a central practical constraint.

Finally, we emphasize that this growth in the number of recovery jump operators is not unique to AutoQEC, but rather reflects a general feature of quantum error correction: enlarging the correctable error set typically requires a correspondingly more intricate recovery structure. The distinctive advantage of AutoQEC is instead architectural: it can realize this protection through always-on engineered dissipation, without continuous monitoring or active feed-forward. The practical question is therefore not whether higher-order AutoQEC is universally simplest to implement, but rather whether the reduction in the required dissipation strength $R$ outweighs the increased number of recovery jump operators in the platform of interest.

\subsection{Complexity Analysis of the LP Feasibility Search}
For a fixed pair of distinct eigenvalues \((h_i,h_j)\), the condition (T2) can be rewritten as a feasibility problem for a vector \(p=(p_i,p_j)\in \mathbb{R}_{\ge 0}^{N_i+N_j}\), where
\begin{align}
A_{ij}^{[\sim c]} \in \mathbb{C}^{|K^{[\sim c]}|\times (N_i+N_j)} .
\end{align}
Thus, for a fixed pair \((i,j)\), the linear program has \(N_i+N_j\) decision variables. In the general complex formulation, the constraints
\begin{align}
\mathrm{Re}\!\left(A_{ij}^{[\sim c]}p\right)=0,
\qquad
\mathrm{Im}\!\left(A_{ij}^{[\sim c]}p\right)=0
\end{align}
contribute up to \(2|K^{[\sim c]}|\) scalar equalities, while the normalization of \(p_i\) and \(p_j\) adds two more constraints. Therefore, the LP size for a fixed pair is controlled by \(N_i+N_j\) and \(|K^{[\sim c]}|\). The overall search then requires checking candidate pairs of distinct eigenvalues of \(\hat H\); if \(d\) denotes the number of distinct eigenvalues of \(\hat H\), this amounts to at most \(O(d^2)\) feasibility problems.

However, although each LP is polynomial in these input dimensions, the input itself can grow rapidly with system size and with the AutoQEC order \(c\). Indeed, since
\begin{align}
K^{[\sim c]}=\{\hat E_a^\dagger \hat E_b \mid \hat E_a,\hat E_b \in E^{[\sim c]}\},
\end{align}
one has the worst-case bound
\begin{align}
|K^{[\sim c]}| \le |E^{[\sim c]}|^2,
\end{align}
and \(|E^{[\sim c]}|\) may itself grow rapidly under the recursive construction of higher-order error sets as discussed in the previous subsection. Accordingly, we do not claim generic worst-case efficiency. Rather, the LP formulation should be viewed as a systematic and explicit method for checking (T2), which is computationally practical for modest system sizes or highly structured models.

We also emphasize that this issue is not unique to AutoQEC, but reflects a broader feature of quantum error correction. In standard QEC as well, verifying exact Knill--Laflamme conditions over a large error set can become computationally demanding. In this sense, the LP complexity reflects the intrinsic complexity of the underlying QEC task, rather than a bottleneck specific to our scheme. In special cases, such as those covered by Corollary~1, the verification can be substantially simplified and the LP search can even be avoided.

\end{document}